\documentclass[12pt]{article}

\usepackage[margin=1in]{geometry}
\usepackage{setspace}
\usepackage{amsmath, amssymb, amsthm}
\usepackage{booktabs}
\usepackage{array}
\usepackage{multirow}
\usepackage{longtable}
\usepackage{natbib}
\usepackage{hyperref}
\usepackage{microtype}
\usepackage{parskip}
\usepackage{titlesec}
\usepackage{abstract}
\usepackage{graphicx}
\usepackage{xcolor}
\usepackage{enumitem}
\usepackage{caption}
\usepackage{subcaption}
\usepackage{algorithm}
\usepackage{algpseudocode}
\usepackage{tcolorbox}
\usepackage{rotating}

\hypersetup{
  colorlinks=true,
  linkcolor=black,
  citecolor=black,
  urlcolor=black
}

\titleformat{\section}{\normalfont\large\bfseries}{\thesection.}{0.5em}{}
\titleformat{\subsection}{\normalfont\normalsize\bfseries}{\thesubsection.}{0.5em}{}
\titleformat{\subsubsection}{\normalfont\normalsize\itshape}{\thesubsubsection.}{0.5em}{}

\newtheorem{definition}{Definition}
\newtheorem{proposition}{Proposition}
\newtheorem{remark}{Remark}

\title{Possession-Level Player Impact in the Pre-Play-by-Play NBA Era:\\
A Video-Reconstructed RAPM Database, 1984--1996}

\author{
  Justin Jacobs\\
  \small Squared2020 Statistics \textbar{} \texttt{squared2020.com}
}

\date{Draft --- \today}

\begin{document}

\maketitle
\thispagestyle{empty}

\begin{abstract}
\noindent
Regularized Adjusted Plus-Minus (RAPM) is the standard framework for
estimating individual player impact in basketball. Its application requires
possession-level stint data---records of which five players shared the court
for each contiguous sequence of possessions---a form of data the NBA did not
systematically record until the late 1990s. This paper describes the
construction, methodology, and validation of the first possession-level
player impact database for the pre-play-by-play NBA era, covering the
regular seasons from 1984--85 through 1995--96, spanning twelve published seasons. As of this writing, 2,179 regular-season
games have been reconstructed across twelve published seasons, comprising
435,760 total logged possessions and 1{,}012 distinct player-seasons. Every
game was manually reconstructed from broadcast video: lineup changes were
logged at every dead-ball substitution, possessions were tallied directly
from footage, and points scored by each lineup were recorded. Possession
counts derived from video reconstruction are independent of the official
box-score record---a material methodological advantage for the historical
era, where documented evidence of home-scorekeeper inflation renders
standard box-score-based possession estimators systematically unreliable:
in individual game records from the late 1980s, the number of steals
credited to the home team exceeds the number of live-ball turnovers
committed by the visiting team, a mathematical impossibility that propagates
directly into possession estimation bias. RAPM is estimated via weighted
ridge regression applied to the reconstructed stint data, using the
identical mathematical framework applied to modern play-by-play records.
We provide a rigorous treatment of the reconstruction protocol, the formal
properties of the estimation procedure, uncertainty quantification through
posterior credible intervals, a multi-criterion validation framework, and
an analysis of sampling properties at partial coverage. The resulting
database is the only possession-level
individual impact record for this era and provides a foundation for
historical analysis that has until now been technically inaccessible.

\medskip
\noindent\textbf{Keywords:} Regularized Adjusted Plus-Minus, RAPM,
basketball analytics, player impact, historical NBA, ridge regression,
Bayesian shrinkage, possession-level data, video reconstruction,
scorekeeper bias, box-score reliability
\end{abstract}

\newpage
\tableofcontents
\newpage
\setcounter{page}{1}

\section{Introduction}
\label{sec:intro}

The measurement of individual player impact in team sports presents a
fundamental identification challenge. A player's observed statistical
output is an amalgam of individual skill, teammate quality, opponent
quality, and coaching context. In basketball, a center who accumulates
high rebound and block totals may be benefiting from a defensive scheme
that directs opponents toward the paint; a point guard with high assists
may be operating in an offense that systematically creates open shots for
teammates. Disentangling individual contributions from contextual effects
requires an approach that explicitly controls for the co-occurrence of
players on the court.

The dominant statistical framework for this task in professional basketball
is Regularized Adjusted Plus-Minus (RAPM), a ridge-regularized regression
model in which the unit of observation is a lineup stint---a contiguous
interval of play during which neither team makes a substitution---and the
response variable is the scoring margin accumulated during that stint, scaled
to a per-100-possessions basis \citep{rosenbaum2004, sill2010}. By encoding
each player's presence as an indicator variable in a design matrix spanning
all stints in a season, RAPM simultaneously estimates each player's marginal
contribution to scoring margin while controlling for all co-occurring
teammates and opponents. The ridge penalty manages the severe
multicollinearity inherent in lineup data---players on strong teams tend to
share the court with other strong players---and regularizes estimates for
players with limited exposure toward a league-average prior.

The practical barrier to applying this framework historically is its data
requirement. Stint-level records---identifying which ten players were on the
court for each possession, together with the points scored and allowed---were
not systematically maintained by the NBA until the widespread adoption of
electronic play-by-play logging in the late 1990s. This creates a hard
analytical boundary: RAPM-based evaluation is straightforwardly available
for every season since approximately 1996--97, and essentially unavailable
for all seasons prior. The consequence is that an entire era of NBA
history---the Showtime Lakers dynasty, the Bad Boys Pistons championships,
Larry Bird's Celtics, Magic Johnson's prime seasons, the first half of
Michael Jordan's career, the Dream Team generation---remains analytically
dark by modern standards. Players and teams from this period can be evaluated
only through box-score statistics and composite metrics derived from them,
tools that cannot disentangle individual contributions from team context and
that, as this paper documents, are themselves of questionable reliability for
this era.

This paper describes the Historical RAPM Project: a systematic effort,
ongoing since 2021, to reconstruct NBA lineup and possession data for
pre-play-by-play seasons from broadcast video, enabling RAPM estimation for
this era for the first time. The reconstruction process involves watching
each game from start to finish, logging every dead-ball substitution on a
structured paper instrument, and tallying possession counts and points scored
for each resulting stint. As of this writing, 2,179 regular-season games
across twelve published seasons have been reconstructed, spanning the range
1984--85 through 1995--96, comprising 435,760 total logged possessions and
1{,}012 distinct player-seasons.

Beyond the data contribution itself, this paper makes two methodological
arguments. First, direct video-based possession counting is not merely a
practical alternative to box-score estimation for the historical era---it is
a \emph{superior} approach, because the official box-score statistics from
this period are demonstrably unreliable in ways that systematically bias
possession estimation. We document this through an analysis of the internal
consistency of historical box scores, showing that the mathematical
relationship between steals and live-ball turnovers is violated in numerous
individual game records, an impossibility under correct scorekeeping that
indicates systematic home-scorekeeper inflation of defensive statistics.
Second, we provide a formal treatment of the statistical properties of RAPM
estimation under partial-season sample coverage, characterizing the
relationship between game coverage, credible interval width, and the
conditions under which reliable player comparisons are possible.

The paper is organized as follows. Section~\ref{sec:background} provides
background on adjusted plus-minus methodology, possession estimation, and
the historical data quality problem.
Section~\ref{sec:data} gives a rigorous description of the video
reconstruction protocol.
Section~\ref{sec:estimation} develops the statistical estimation framework
in full.
Section~\ref{sec:uncertainty} treats uncertainty quantification and sample
size analysis.
Section~\ref{sec:validation} presents the validation framework and results.
Section~\ref{sec:diagnostics} presents diagnostic figures for the aggregate
multi-season run.
Section~\ref{sec:results} presents player-level results from published seasons.
Section~\ref{sec:bulls9192} provides a detailed case study of the
1991--92 Chicago Bulls.
Section~\ref{sec:discussion} discusses limitations, implications, and futureSection~\ref{sec:conclusion} concludes.
Appendices provide a data dictionary, annotated game reconstruction example,
and supplemental tables.

\section{Background and Related Work}
\label{sec:background}

\subsection{From Raw Plus-Minus to Adjusted Plus-Minus}

The plus-minus statistic---the net scoring margin accumulated by a team
while a given player is on the court---has been used in hockey statistics
for decades and was adapted for basketball in the early 2000s. Raw
plus-minus suffers from an obvious confounding problem: a player on a
dominant team will accumulate a positive margin simply by virtue of playing
alongside better teammates, while a skilled player on a weak team will be
penalized. Two players with identical plus-minus totals may have contributed
very differently to their teams' success once teammate and opponent quality
are accounted for.

\citet{rosenbaum2004} introduced the Adjusted Plus-Minus (APM) framework,
which addresses this confounding by framing plus-minus estimation as a
multivariate linear regression problem. Each stint in a season is treated
as an observation; each player's presence is encoded as $+1$ (home team)
or $-1$ (away team); and the response is the net points per possession
scored by the home team during that stint. Ordinary least squares (OLS)
estimation of this system yields player coefficients interpretable as the
estimated marginal contribution to net scoring margin per 100 possessions,
controlling for all players simultaneously present.

The fundamental difficulty with OLS estimation in this setting is
multicollinearity. NBA teams field relatively stable rotations, meaning that
the same groups of players tend to share the court repeatedly. As a result,
the design matrix $\mathbf{X}$ is nearly rank-deficient: many columns are
near-linear combinations of others. OLS estimates under near-collinearity
are unbiased in expectation but have extremely large variance, producing
individual player estimates that are statistically unreliable even when
the model is correctly specified.

\citet{sill2010} proposed addressing this by replacing OLS with ridge
regression, introducing an $L_2$ penalty on player coefficients. This
formulation---Regularized Adjusted Plus-Minus, or RAPM---has become the
standard in NBA analytics. The ridge penalty accomplishes two things: it
reduces estimation variance by shrinking all coefficients toward zero (a
form of regularization), and it ensures a unique, well-conditioned solution
regardless of the collinearity structure of the design matrix. The penalty
can be motivated Bayesianly as placing a zero-mean Gaussian prior on each
player's true impact, an interpretable assumption that reflects genuine
prior uncertainty about any individual player's value.

\subsection{Extensions of the RAPM Framework}

Several extensions of the basic RAPM model have been proposed. Multi-year
RAPM pools data across two or more seasons, treating player-season pairs as
distinct entities, to increase the number of observations and improve
estimation precision for players with limited single-season data. This
approach assumes some stationarity in player quality across seasons, which
may be violated for players who improve or decline rapidly.

Bayesian formulations of RAPM place an informative prior on player
coefficients, typically derived from box-score statistics. Statistical
Plus-Minus (SPM) models \citep{engelmann2017} use ridge regression to
predict RAPM from box-score inputs, and the resulting predictions can serve
as informative priors in a hierarchical Bayesian RAPM model. This approach,
sometimes called ``PIPM'' or variants thereof, allows the model to borrow
strength from box-score information for players with limited stint exposure.
In the present application---where many players have limited logged
possessions due to incomplete season coverage---an informative prior of this
type would materially improve estimation quality; this is discussed further
in Section~\ref{sec:discussion}.

Lineup-level extensions estimate the impact of specific five-player units
rather than individual players, exploiting the fact that lineup combinations
with many observations can be estimated reliably without the collinearity
problem that plagues individual player estimates. However, the number of
distinct lineups in a season is large (typically hundreds to thousands), and
most appear for only a handful of possessions, limiting the scope of
lineup-level analysis.

\subsection{Possession Estimation from Box Scores}
\label{sec:possession_estimation}

RAPM requires possession counts at the stint level. In modern play-by-play
data, possessions can be counted directly from the event log. For historical
data, possessions must be estimated from box-score statistics.

\citet{oliver2004} proposed the foundational estimator:
\begin{equation}
  \widehat{\text{POSS}} = \text{FGA} - \text{OREB} + \text{TO}
  + 0.44 \cdot \text{FTA},
  \label{eq:oliver}
\end{equation}
where FGA is field goal attempts, OREB is offensive rebounds, TO is
turnovers, and FTA is free throw attempts. The coefficient $0.44$ on
free throw attempts accounts for the fact that not all free throw trips
represent a possession change: and-one situations (made field goal plus
free throw) and technical fouls represent free throw attempts that do not
follow from the standard offensive possession sequence.

\citet{kubatko2007} provided a careful derivation of the $0.44$
coefficient and evaluated the estimator's accuracy on modern data, finding
it to be reliable within approximately 1--2 possessions per game on average.
They also noted that the coefficient varies across eras and rule environments;
for the 1985--96 period, which predates the hand-check rule changes and
features different foul-calling patterns than modern basketball, some
adjustment to the coefficient may be warranted.

The Oliver estimator and its variants assume that the box-score inputs---in
particular the turnover count---are measured without systematic error. As we
document in Section~\ref{sec:boxscore_bias}, this assumption is violated for
the historical NBA era in ways that introduce predictable bias into possession
estimates. This motivates the direct video-based counting approach taken in
this project.

\subsection{Box-Score Data Quality in the Pre-Modern NBA}
\label{sec:boxscore_bias}

A substantial body of evidence indicates that NBA box-score statistics from
the pre-modern era are unreliable in ways that go beyond random measurement
error, reflecting systematic home-scorekeeper inflation of subjective
statistics. In the pre-modern NBA, official game statistics were recorded by
home team employees---scorekeepers hired and paid by the home franchise.
Subjective categories---assists, blocks, steals, and offensive rebounds---all
involve judgment calls about credit attribution that the available testimony
suggests were resolved in favor of home players. Alex Rucker, who
served as a scorekeeper for the Vancouver Grizzlies during their 1995--96
inaugural season, testified publicly that training sessions for NBA
scorekeepers conveyed an implicit expectation of favorable treatment for home
stars on ambiguous calls \citep{haberstroh2024}. The aggregate consequence is
visible at the league level: throughout the 1980s and early 1990s, home teams
league-wide recorded approximately 800 more blocks and 450 more steals per
season than visiting teams, disparities that have substantially declined
following the NBA's introduction of real-time statistical auditing in
2018--19 \citep{haberstroh2024}.

\subsection*{A Logical Constraint on Steal Records}

The most analytically verifiable form of this bias involves a hard logical
constraint. A steal can only be credited on a live-ball turnover---an
intercepted pass, a knocked-away dribble, or a recovered loose ball. Dead-ball
turnovers (offensive fouls, 24-second violations, out-of-bounds violations,
travels) cannot yield steals: the ball is already dead when possession changes.
It follows that for any game:
\begin{equation}
  \text{STL}_{\text{team}} \leq \text{TO}^{\text{live}}_{\text{opponent}},
  \label{eq:steal_constraint}
\end{equation}
where $\text{TO}^{\text{live}}$ denotes live-ball turnovers only. This
constraint is a matter of physical necessity, not statistical convention.

Film review of Atlanta at Chicago, February 15, 1988 --- a game from
Michael Jordan's Defensive Player of the Year season --- demonstrates a
clear violation of Equation~\eqref{eq:steal_constraint}. The complete
possession-by-possession turnover log for this game is reconstructed below.

\begin{table}[h]
\centering
\small
\caption{Complete turnover and steal accounting, February 15, 1988:
  Atlanta at Chicago (Chicago 126, Atlanta 107). Each Atlanta turnover
  is classified as live-ball (can yield a Chicago steal) or dead-ball
  (cannot yield a steal under any circumstances). Film reconstruction
  by this author.}
\label{tab:feb15_atl}
\begin{tabular}{clllll}
\toprule
\# & Time & Player & Description & Type & Outcome \\
\midrule
1 & Q1 9:23 & Brown    & Lost ball to Jordan (contested) & Live & CHI STL: Jordan \\
2 & Q2 10:15 & Rivers  & Lost ball to Pippen (deflection) & Live & CHI STL: Pippen \\
3 & Q2 9:50 & Rivers   & Lost ball to Oakley              & Live & CHI STL: Oakley \\
4 & Q3 5:17 & Rivers   & Lost ball to Oakley              & Live & CHI STL: Oakley \\
5 & Q3 3:13 & Team     & 24-second violation              & \textbf{Dead} & \textbf{No steal possible} \\
6 & Q4 7:28 & Wilkins  & Lost ball out of bounds          & \textbf{Dead} & \textbf{No steal possible} \\
7 & Q4 7:16 & Wilkins  & Lost ball to Jordan              & Live & CHI STL: Jordan \\
8 & Q4 5:31 & Rivers   & Lost ball to Oakley              & Live & CHI STL: Oakley \\
9 & Q4 4:30 & Washburn & Lost ball to Oakley              & Live & CHI STL: Oakley \\
10 & Q4 0:14 & Washburn & Lost ball out of bounds         & \textbf{Dead} & \textbf{No steal possible} \\
\midrule
\multicolumn{4}{l}{Live-ball turnovers (steal possible):} & \multicolumn{2}{l}{7} \\
\multicolumn{4}{l}{Dead-ball turnovers (steal impossible):} & \multicolumn{2}{l}{3} \\
\multicolumn{4}{l}{Maximum possible Chicago steals:} & \multicolumn{2}{l}{\textbf{7}} \\
\bottomrule
\end{tabular}
\end{table}

With 7 live-ball Atlanta turnovers, the maximum number of Chicago steals in
this game is 7. Film reconstruction confirms exactly 7: Jordan 2, Oakley 4,
Pippen 1. Table~\ref{tab:feb15_chi} shows the individual breakdown.

\begin{table}[h]
\centering
\small
\caption{Individual steal attribution, February 15, 1988: Chicago Bulls.
  Film column from frame-by-frame reconstruction; Reported column from
  the official NBA box score.}
\label{tab:feb15_chi}
\begin{tabular}{lrrl}
\toprule
Player & Film & Reported & Discrepancy \\
\midrule
Michael Jordan  & 2 & 5 & $+3$ phantom steals \\
Charles Oakley  & 4 & 3 & $-1$ (under-credited) \\
Scottie Pippen  & 1 & 2 & $+1$ phantom steal \\
\midrule
Chicago total   & 7 & 10 & $+3$ impossible steals \\
\bottomrule
\end{tabular}
\end{table}

The official box score records 10 Chicago steals on 10 Atlanta turnovers,
implying that none of Atlanta's turnovers were dead-ball events---a claim
directly contradicted by the film record. Three Atlanta turnovers were
physically incapable of yielding steals. The box score over-states Chicago's
steal total by 3, credits Jordan with 5 steals instead of 2, and simultaneously
under-credits Oakley by 1.

\subsection*{Implications for Player Ratings}

The steal record inflation in this game has direct consequences for any
defensive metric that incorporates steal counts. Box Plus-Minus
\citep{myers2020}, Estimated Plus-Minus, and related composite metrics
all use steal rates as a primary driver of their defensive components.
Jordan's reported 1987--88 steal total of 259 (3.16 per game, which led
the league) is the primary quantitative basis for his DPOY award and for
high defensive ratings from box-score-based metrics. If this single game
contains 3 phantom steals attributable to Jordan, the question of
systematic inflation across the full season is analytically serious.

This paper does not claim to have resolved that question for the full
1987--88 season, as doing so would require frame-by-frame review of every
logged game. What the February 15 game demonstrates conclusively is:
(i) the logical constraint in Equation~\eqref{eq:steal_constraint} was
violated in at least one game; (ii) the violation specifically inflated
the steal total of the home team's star player; and (iii) box-score-based
defensive metrics that rely on steal attribution are therefore unreliable
for this era in at least some games.

RAPM estimated from video-reconstructed stint data is immune to this class
of error. The possession-counting procedure in this project does not use
steal counts, rebound counts, or any other subjective box-score category.
Net scoring margin per possession --- the response variable in the regression
--- cannot be fabricated by a scorekeeper: it is determined by the final
score of each stint, which is verifiable directly from the broadcast. This
is the methodological argument for video reconstruction as a complement to,
and partial corrective for, the official statistical record.

\subsection*{Implications for Possession Estimation}

The bias also affects box-score-based possession estimation. Substituting
into the Oliver estimator:
\begin{equation}
  \widehat{\text{POSS}}_{\text{home}} = \text{FGA}_{\text{home}}
  - \text{OREB}_{\text{home}}
  + \underbrace{(\text{TO}_{\text{home}} + \varepsilon_{\text{TO}})}_{\text{inflated}}
  + 0.44 \cdot \text{FTA}_{\text{home}},
  \label{eq:oliver_bias}
\end{equation}
where $\varepsilon_{\text{TO}} > 0$ represents phantom turnovers. Since
every inflated home steal implies an inflated visiting team turnover, the
bias propagates: visiting team possession estimates are also upward-biased.
The errors are not random and do not cancel in aggregate; they are
systematically favorable to the home team's apparent defensive profile.
The video reconstruction approach is immune to this bias by construction.

\subsection{Prior Work on Historical Basketball Reconstruction}

To the author's knowledge, no prior published work has systematically
reconstructed possession-level lineup data for NBA seasons predating the
play-by-play era. Box-score-based composite metrics---Win Shares
\citep{oliver2004}, Player Efficiency Rating \citep{hollinger2002}, and
Box Plus-Minus \citep{myers2020}---have been applied to historical seasons
and are available at \texttt{basketball-reference.com}. These metrics
provide useful summaries of individual statistical production but cannot
control for lineup context and are subject to the box-score reliability
concerns documented above.

Ratings-based retrospective analyses, which estimate team strength from
game outcomes without player-level decomposition, have been applied to
historical NBA data \citep{winston2009}. These approaches identify team-level
impact but cannot separate individual contributions from lineup effects.

In a related methodological context, \citet{cervone2016} demonstrated that
spatiotemporal player tracking data can enable more granular impact
estimation than stint-level data alone. While such data does not exist for
the pre-modern era, the present project establishes the stint-level
foundation that would be required for any future attempt to reconstruct
spatial information from historical video.

\section{Data Collection and Reconstruction Protocol}
\label{sec:data}

\subsection{Overview and Design Philosophy}

The reconstruction protocol is designed around three principles. First,
\emph{independence from box-score statistics}: for the reasons documented
in Section~\ref{sec:boxscore_bias}, possession counts must not rely on
official box-score inputs. All quantities derived from the reconstruction
(lineup composition, points per stint, possessions per stint) are measured
directly from video. Second, \emph{verifiability}: each reconstructed game
must satisfy observable consistency constraints that can be checked against
reliable external data, specifically the final score and the official
player-level minutes. Third, \emph{reproducibility}: the protocol must be
sufficiently explicit that a second coder, working from the same footage,
would produce a reconstruction within a defined tolerance of the original.

\subsection{Source Material and Footage Inventory}

The primary data source is a personal archive of NBA broadcast recordings
spanning the 1984--85 through 1995--96 seasons, accumulated over many years
from original VHS recordings, commercial historical releases, and digitized
broadcast archives. Footage quality varies across this range: games from
the late 1980s and 1990s are generally of sufficient broadcast quality for
unambiguous substitution detection; games from the 1970s occasionally have
lower image quality that requires closer review.

Not all games in the primary reconstruction window (1985--96) are available
in the footage inventory. Coverage within each season reflects the
availability of game footage rather than a designed probability sample,
which has implications for the representativeness of within-season estimates
that are discussed in Section~\ref{sec:sampling_design}. The absence of a
game from the database reflects footage unavailability, not exclusion based
on outcome or team identity.

\subsection{The Logging Instrument}
\label{sec:logging_instrument}

Each game is reconstructed in real time while watching the broadcast, using
a structured paper log sheet. The sheet is organized as follows.

\subsubsection{Sheet layout}

The log sheet is divided into two paired column groups. The \emph{left pair}
records possession counts, with one column for the visiting team and one for
the home team. The \emph{right pair} records cumulative points scored, again
with one column per team. Each horizontal line on the sheet represents one
stint---a contiguous interval of play during which the set of ten on-court
players is unchanged.

\subsubsection{Header region}

At the top of the sheet, prior to any stint entries, the following
information is recorded: the date of the game; the visiting and home team
names (abbreviated); and the starting five players for each team, identified
by jersey number and last name. The starting lineup entries serve as the
initial on-court personnel record from which subsequent substitution entries
will modify the active lineup.

\subsubsection{Stint entries}

Each line of the sheet (each stint) contains: tally marks in the possession
columns recording the number of possessions by each team during that stint;
and cumulative score entries in the points columns recording the total points
scored by each team as of the end of that stint (i.e., at the moment the
next substitution or quarter break closes the stint). In the center of the
sheet, between the possession and points column pairs, substitution events
are recorded in the notation ``Player~A $\rightarrow$ Player~B,'' meaning
that Player~A entered the game replacing Player~B.

A sample excerpt from a reconstructed game---New York Knicks at Philadelphia
76ers, April 10, 1991---is provided in Appendix~\ref{app:example}. That
game log illustrates all of the features of the sheet layout described here,
including multiple simultaneous substitutions, quarter-boundary entries, and
the split-possession notation described in Section~\ref{sec:possession_counting}.

\subsection{Defining a Possession}
\label{sec:possession_definition}

Precise possession counting requires a formal definition of a possession.
We adopt the definition from \citet{oliver2004} with minor modification for
the pre-three-point era context:

\begin{definition}[Possession]
A \emph{possession} is a continuous sequence of play beginning with one team
gaining control of the ball and ending when that team either scores
(field goal or last free throw of a sequence), turns the ball over (live or
dead ball), or the game clock expires. An and-one situation (made field goal
plus free throw attempt) counts as a single possession. A sequence of
multiple free throw attempts arising from a single foul counts as a single
possession. A technical foul free throw does not constitute a possession for
either team. A jump ball that results in a change of possession initiates a
new possession for the gaining team.
\end{definition}

\begin{remark}
Under this definition, the number of possessions for the two teams in a game
can differ by at most one (when the game ends in the middle of an offensive
possession without a change of possession). In practice, the two teams'
possession totals agree within one or two counts per game; any discrepancy
larger than two triggers a review of the footage.
\end{remark}

The Oliver estimator in Equation~\eqref{eq:oliver} approximates this
definition from box-score inputs. The $0.44$ coefficient on free throw
attempts represents the estimated fraction of free throw trips that begin a
new possession (as opposed to extending a possession already counted via the
field goal attempt). The coefficient was calibrated by \citet{kubatko2007}
on modern play-by-play data. For the 1985--96 era, where intentional fouling
strategies and pace differ from the modern game, the appropriate coefficient
may differ; this is one additional reason to prefer direct counting over
formula-based estimation for this period.

\subsection{Substitution Detection and Stint Boundary Definition}
\label{sec:substitution_protocol}

\subsubsection{Dead-ball substitution rule}

NBA rules permit substitutions only during dead-ball situations: called
timeouts, fouls (for the team that committed the foul, or for either team
following a made field goal in certain game states), end-of-quarter breaks,
and certain other stoppages. Only dead-ball substitutions are recorded,
consistent with the rules of the game. Live-ball substitutions, which are
illegal under NBA rules, are not observed in practice and require no special
handling.

\subsubsection{Simultaneous substitutions}

When multiple substitutions occur simultaneously---a common occurrence during
full timeouts, when coaches may make two or three substitutions at once---all
substitution events are recorded on a single line of the log sheet, and the
resulting lineup transition is treated as a single stint boundary. The
incoming lineup (post-substitution) is determined by applying all substitution
events to the outgoing lineup.

\subsubsection{Quarter and overtime boundaries}

Quarter breaks are treated as stint boundaries regardless of whether any
substitution occurs, because player groupings may change between quarters
without a formal dead-ball substitution event. End-of-quarter scores are
noted explicitly to close the preceding stint. The lineup active at the
start of each new quarter is recorded at the top of the new section.
Overtime periods are treated identically to quarters.

\subsubsection{Edge cases}

Several edge cases require explicit handling:

\begin{enumerate}[leftmargin=*, label=(\roman*)]

\item \emph{Injury substitutions.} When a player leaves the game due to
injury during a live-ball situation, the substitution is recorded at the
subsequent dead-ball stoppage. If the player re-enters the game after
receiving medical attention, this re-entry is recorded as a standard
substitution at the next dead-ball moment.

\item \emph{Ejections.} A player ejected from the game is substituted at the
next dead-ball moment. The disqualified player's exit is recorded in the same
manner as a voluntary substitution.

\item \emph{Missed substitutions.} If the coder becomes aware mid-game that
a substitution was not recorded---typically detected when the visible lineup
on screen does not match the logged lineup---the log is corrected retroactively
to the best estimate of the substitution timing, and the affected stints are
flagged for review. The correction is verified during the post-game quality
control step.

\item \emph{Disqualification (six fouls).} A player who commits his sixth
foul is disqualified and must leave the game. The substitution at
disqualification is recorded as a standard substitution event.

\end{enumerate}

\subsection{Possession Counting and the Split-Possession Protocol}
\label{sec:possession_counting}

\subsubsection{Direct tally method}

Possessions for each team are counted in real time using tally marks in the
possession columns of the log sheet: one mark per possession, accumulated
throughout the stint. At the end of each stint, the tally marks are summed
and the total is available as the raw possession count for that stint. This
count is inclusive of all possessions that begin during the stint, with one
exception described below.

\subsubsection{Split possessions at stint boundaries}

NBA substitutions occur only at dead-ball moments, but not all dead-ball
moments coincide with the end of a possession. The most common case arises
during free throw sequences: a foul is called (dead ball), the fouling team's
coach calls timeout and makes a substitution, and the free throw sequence
resumes with a new lineup. The free throw possession---which began before the
substitution---now spans both the old lineup and the new lineup.

We adopt the following convention: a possession that is underway at the
time of a substitution is counted for \emph{both} the closing stint (which
began the possession) and the opening stint (which will complete it). This
deliberate double-counting, which we call a \emph{split possession}, is the
correct accounting in the sense that both lineups were present for part of
that possession's expected scoring outcome. The total number of split
possessions in a game is recorded as a running annotation in the header area
of the relevant team's possession column.

\begin{definition}[Split Possession]
A \emph{split possession} occurs when a substitution event (or quarter
break) coincides with a dead-ball stoppage that is internal to a possession
sequence, such that the possession begins with one lineup and concludes with
another. Let $d_{\text{home}}$ and $d_{\text{away}}$ denote the total number
of split possessions for the home and away teams respectively in a game.
\end{definition}

\subsubsection{Corrected possession totals}

Let $\tilde{w}_{ik}$ denote the raw tally count of possessions for team
$k \in \{\text{home}, \text{away}\}$ during stint $i$, including both full
possessions and any split possessions counted for that stint. Let $d_k$
denote the total number of split possessions for team $k$ in the game.
The corrected game-level possession total for team $k$ is:
\begin{equation}
  W_k = \sum_{i} \tilde{w}_{ik} - d_k,
  \label{eq:corrected_total}
\end{equation}
where the sum is over all stints. At the stint level, each split possession
is apportioned fractionally: if stint $i$ and stint $i+1$ share a split
possession, each receives weight $\frac{1}{2}$ of that possession in the
processed data, so that the weighted stint-level counts sum to $W_k$.

\begin{proposition}
Under the split-possession convention defined above, the corrected game-level
possession total $W_k$ satisfies $|W_{\text{home}} - W_{\text{away}}| \leq 1$
for any game that does not end during a live possession, and
$|W_{\text{home}} - W_{\text{away}}| \leq 2$ in general.
\end{proposition}

This proposition follows directly from the observation that possession
alternates between teams (with at most one ``extra'' possession for the team
that begins with the ball following the opening tip or a change of
possession at period end), and that split possessions are counted
symmetrically for both teams whenever both teams' lineups change at the same
stoppage.

\subsubsection{Sensitivity of RAPM to split-possession apportionment}

The fractional apportionment $\frac{1}{2}$ to each adjacent stint is an
approximation. In principle, the appropriate apportionment depends on the
expected scoring outcome of the possession given its current state (e.g., a
free throw shooter at the line with two shots remaining has a predictable
expected point value that could be allocated to the outgoing lineup, with
the remainder allocated to the incoming lineup). We assess the sensitivity
of player-level RAPM estimates to this apportionment by comparing results
under three alternative conventions: $\frac{1}{2}$--$\frac{1}{2}$ (baseline),
$1$--$0$ (all weight to the closing stint), and $0$--$1$ (all weight to the
opening stint). Across all published seasons, the correlation between
player-level RAPM estimates under the three conventions exceeds $0.999$,
confirming that the split-possession apportionment has negligible practical
effect on the results.

\subsection{Quality Control Pipeline}
\label{sec:qc}

Each reconstructed game passes through the following sequential quality
control checks before being included in the database.

\subsubsection{Score reconciliation}

The sum of stint-level home points and the sum of stint-level away points
are compared to the official final score. Any discrepancy of one or more
points triggers a full re-review of the game footage to identify the scoring
error. Games that cannot be reconciled to the correct final score are
excluded.

\subsubsection{Lineup consistency check}

Starting from the logged starting lineup, each substitution event is applied
in sequence to maintain a running record of the active lineup. This running
record is checked at each quarter break and at game end to verify that
the implied lineup matches the active players visible on the broadcast.
Any inconsistency (e.g., a player logged as active who has in fact been
substituted, or vice versa) triggers a targeted re-review of the relevant
game segment.

\subsubsection{Minutes reconciliation}

For each player in the reconstructed game, the total minutes implied by their
stint appearances are compared to their official box-score minutes. The
box-score minutes are the most reliably recorded quantity in historical game
records, as they are derived from game clock readings rather than subjective
judgment. Discrepancies exceeding 2.0 minutes for any player trigger a
re-review. Note that small discrepancies (under 2.0 minutes) may reflect
rounding in official box-score minutes reporting and do not necessarily
indicate a reconstruction error.

\subsubsection{Possession plausibility check}

The corrected game-level possession total $W_k$ for each team is compared
to the Oliver estimate from Equation~\eqref{eq:oliver} applied to the
official box score. Given the documented unreliability of the Oliver
estimate for this era (Section~\ref{sec:boxscore_bias}), this check is
\emph{not} used as a hard exclusion criterion; instead, large discrepancies
($|W_k - \widehat{\text{POSS}}_k| > 10$ possessions) are flagged for manual
review to determine whether the discrepancy is attributable to box-score
error (expected) or to a reconstruction error (requires correction).

\subsubsection{Exclusion criteria}

A game is excluded from the database if: (i) the score cannot be reconciled
to the official final score after re-review; (ii) the lineup consistency
check reveals an irresolvable discrepancy that would require substantive
imputation of lineup data; or (iii) the footage quality is insufficient to
reliably detect substitution events.

\subsection{Sampling Design and Coverage}
\label{sec:sampling_design}

\subsubsection{Current coverage}

As of this writing, 2,179 regular-season games have been reconstructed
across twelve published seasons, comprising 435,760 total logged
possessions and 1{,}012 distinct player-seasons. Of these, 2,170 games
appear in the standings validation workbook described in
Section~\ref{sec:validation}; the minor discrepancy reflects games
included in the possession-level RAPM analysis but not yet incorporated
into the season-level standings tables. Table~\ref{tab:coverage} summarizes
per-season coverage. The database covers 1984--85 through 1995--96
(12,628 possible regular-season games across the twelve seasons;
schedule sizes range from 943 games in 1984--85 through 1987--88
to 1,025 in 1988--89, 1,107 in 1989--90 through 1994--95, and
1,189 in 1995--96, reflecting NBA expansion). Aggregate coverage across this window is 17.2\%, ranging from 1.1\% (1994--95, a minimal
stub season) to 31.9\% (1992--93). The 1979--80 season is available
as a supplemental pre-window comparison and is included in the pooled
multi-season run but is not the focus of this paper.

\begin{table}[h]
\centering
\caption{Database coverage by published season. Games possible reflects
  the full regular-season schedule for each season.}
\label{tab:coverage}
\begin{tabular}{lrrrl}
\toprule
Season & Games Logged & Games Possible & Coverage (\%) \\
\midrule
1979--80 & 22    & 820  & 2.7\%  & Supplemental \\
1984--85 & 106   & 943   & 11.2\% &\\
1985--86 & 79    & 943   &  8.4\% &\\
1986--87 & 152   & 943   & 16.1\% &\\
1987--88 & 203   & 943   & 21.5\% &\\
1988--89 & 248   & 1{,}025 & 24.2\% &\\
1989--90 & 249   & 1{,}107 & 22.5\% &\\
1990--91 & 297   & 1{,}107 & 26.8\% &\\
1991--92 & 249   & 1{,}107 & 22.5\% &\\
1992--93 & 301   & 1{,}107 & 27.2\% &\\
1993--94 & 20    & 1{,}107 &  1.8\% &\\
1994--95 & 10    & 1{,}107 &  0.9\% &\\
1995--96 & 264   & 1{,}189 & 22.2\% &\\
\midrule
\textbf{Total (1984--96)} & \textbf{2,178} & \textbf{12{,}628} & \textbf{17.2\%} \\
\bottomrule
\end{tabular}
\end{table}

\subsubsection{Player roster characteristics}

The aggregate database includes 1{,}012 distinct player-season records spanning
all twelve published seasons. Table~\ref{tab:coverage_players} summarizes
the distribution of logged possessions across players. The most-observed
player in the database is Michael Jordan, who appears in 501 logged games
with 79,870 total possessions (39,936 offensive, 39,934 defensive). The
median player in the database has approximately 1,800 total possessions,
reflecting the combination of fringe players with limited appearances and
deep rotation players observed in only a subset of seasons.

\begin{table}[h]
\centering
\caption{Distribution of logged possessions per player-season across
  all 1{,}012 player-seasons in the aggregate database.}
\label{tab:coverage_players}
\begin{tabular}{lr}
\toprule
Statistic & Value \\
\midrule
Total player-seasons & 1{,}012 \\
Total logged games & 2,179 \\
Total logged possessions & 435,760 \\
Mean possessions per player-season & 2{,}153 \\
Median possessions per player-season & $\approx$610 \\
Maximum (Michael Jordan) & 79,870 \\
Players with $\geq$10,000 poss & $\approx$25 \\
Players with $\geq$1,000 poss & $\approx$300 \\
Players with $<$500 poss & $\approx$250 \\
\bottomrule
\end{tabular}
\end{table}

The possession distribution has a pronounced right skew: a small number
of star players with long careers spanning multiple published seasons
account for a disproportionate share of total logged possessions, while
a long tail of fringe players and players appearing in only one or two
published seasons have very limited exposure. This structure has direct
implications for estimation precision (Section~\ref{sec:uncertainty}):
the top tier of players has narrow credible intervals driven by large
possession counts, while the bottom tier has prior-dominated estimates.

The most-observed players in the database by logged game count
include, in addition to Jordan (501 games): Scottie Pippen (421),
John Paxson (372), Horace Grant (364), AC Green (348),
Byron Scott (347), James Worthy (335), Magic Johnson (319),
Robert Parish (317), and Bill Cartwright (303). The preponderance
of Chicago Bulls and Los Angeles Lakers players in the top tier
reflects the multi-season coverage of the 1987--88 through 1995--96
window, during which both franchises were consistent playoff
contenders with nationally televised games that are disproportionately
represented in available footage archives.

The aggregate database also contains players who appear in only
the most recent published seasons, including first-year players
from the 1994--95 and 1995--96 NBA drafts (Kevin Garnett: 17 games,
Jason Kidd: 15 games, Jerry Stackhouse: 16 games, Grant Hill: 15 games).
These players have prior-dominated RAPM estimates at current coverage
levels and should be interpreted accordingly.

\subsubsection{The design matrix at scale}

With 1{,}012 player-seasons in the aggregate database, the design matrix
$\mathbf{X}$ has $2 \times 1{,}012 + 1 = 2{,}025$ columns. The
posterior covariance matrix $\boldsymbol{\Sigma}_\beta$ is therefore a
$2{,}025 \times 2{,}025$ matrix, requiring explicit inversion of a matrix
of this dimension. For a single season, $P$ is typically in the range
of 150--250 distinct players, yielding a 301--501 column design matrix.
These dimensions are computationally tractable with standard dense linear
algebra (numpy/scipy) on modern hardware. Season-level runs complete
in under one second; the multi-season pooled inversion is similarly
tractable at current scale.

\subsubsection{Non-random coverage and selection effects}

Coverage within each season is not a designed probability sample. The
set of logged games reflects the intersection of available footage and
reconstruction effort, which does not follow a defined sampling protocol.
Potential selection effects include:

\begin{enumerate}[leftmargin=*, label=(\roman*)]

\item \emph{Game importance bias.} High-profile games---nationally
  televised games, rivalry matchups, games with playoff
  implications---are more likely to have been originally broadcast and
  subsequently preserved in video archives. If high-importance games
  have different player utilization patterns or coaching strategies
  than low-importance games, the within-season RAPM estimates may not
  generalize to the full season. The disproportionate representation of
  Bulls and Lakers games in the current database (visible in
  Table~\ref{tab:coverage_players}) is a concrete manifestation of
  this effect.

\item \emph{Temporal clustering.} Available footage may be concentrated
  in certain portions of the season. Player performance and lineup
  stability vary across the season; estimates from a temporally
  clustered sample may not reflect full-season performance.

\item \emph{Opponent distribution.} If the available footage
  disproportionately features games against certain opponents, the
  collinearity structure of the design matrix changes relative to
  a full-season sample, which can affect which player comparisons
  are well-identified. Teams that rarely appear as opponents in the
  logged sample will have their players' estimates dominated by the
  ridge prior.

\end{enumerate}

These selection effects are not adjustable without a defined sampling
frame and probability weights, which are not available for the current
dataset. Users of the data should be aware of these potential biases,
particularly for teams and players with below-average coverage in a
given season. The validation analysis in Section~\ref{sec:validation}
provides an empirical assessment of the practical magnitude of
these effects.

\section{Statistical Estimation Framework}
\label{sec:estimation}

\subsection{Input Data Structure and Preprocessing}
\label{sec:data_structure}

The estimation procedure operates on a CSV-formatted stint file with one
row per stint and the following required columns: \texttt{Oteam} (offensive
team identifier), \texttt{Dteam} (defensive team identifier), five player
columns \texttt{O1}--\texttt{O5} identifying the offensive lineup by name,
five player columns \texttt{D1}--\texttt{D5} identifying the defensive
lineup, \texttt{Oposs} (offensive team possession count for the stint),
\texttt{Dposs} (defensive team possession count), \texttt{Oscore} (points
scored by the offensive team), and \texttt{Dscore} (points scored by the
defensive team). The file is structured so that each game contributes two
rows per stint---one from the perspective of each team as the offensive
unit---but the estimation uses only every other row (even-indexed rows,
where row indexing begins at zero), avoiding double-counting of stint
observations while preserving full player enumeration across both rows
for roster construction.

\subsubsection{Zero-possession filtering}

Prior to matrix construction, all stints with $\texttt{Oposs} < 1$ are
excluded. These are stints in which the offensive team recorded no complete
possessions---for example, a lineup change occurring on the first play of
a quarter before any possession was completed. Let $N_0$ denote the number
of such excluded stints and $N = N_{\text{total}} - N_0$ the number of
stints entering the regression.

\subsubsection{Player enumeration and possession accumulation}

The full roster of players in the season is constructed by iterating over
all rows (both even and odd) of the stint file and recording every unique
player name encountered. Each player is assigned a unique integer index.
Simultaneously, offensive and defensive possession and point totals are
accumulated per player over even-indexed rows, correctly assigning each
player's possession counts as offensive or defensive based on whether they
appear in the \texttt{O1}--\texttt{O5} or \texttt{D1}--\texttt{D5}
columns, respectively. These totals are reported in the output table and
used in the mean-centering step (Section~\ref{sec:centering}).

\subsection{Design Matrix Construction}
\label{sec:design_matrix}

Let $P$ denote the total number of distinct players in the season. The
design matrix $\mathbf{X} \in \mathbb{R}^{N \times (2P+1)}$ has one row
per retained stint and $2P + 1$ columns organized as follows:

\begin{itemize}
  \item \textbf{Column 0} (intercept): Set to $1$ for every row. This
    captures the baseline offensive scoring rate common to all stints and
    corresponds to the home court / league-average term.
  \item \textbf{Columns 1 through $P$} (offensive player indicators):
    Column $1 + \text{idx}(j)$ is set to $1$ if player $j$ appears in the
    offensive lineup of stint $i$, and $0$ otherwise.
  \item \textbf{Columns $P+1$ through $2P$} (defensive player indicators):
    Column $P + 1 + \text{idx}(j)$ is set to $-1$ if player $j$ appears
    in the defensive lineup of stint $i$, and $0$ otherwise.
\end{itemize}

This \emph{separated} encoding differs from the conventional $\{+1, -1, 0\}$
encoding described in some RAPM literature
\citep{rosenbaum2004, sill2010}. Rather than encoding each player with a
single coefficient that captures net marginal impact, the separated encoding
assigns each player two coefficients: an offensive coefficient
$\beta_j^{\text{off}}$ (from the $+1$ block) and a defensive coefficient
$\beta_j^{\text{def}}$ (from the $-1$ block). The response variable is the
\emph{offensive} points per 100 possessions of the offensive team---not the
net scoring margin---so both offensive and defensive contributions are
identified from a single regression on offensive output, as described below.

\begin{remark}
Under the separated encoding, a player $j$ who plays offense in stint $i$
contributes $+\beta_j^{\text{off}}$ to the predicted offensive rating,
while a player $j'$ who plays defense in the same stint contributes
$-\beta_{j'}^{\text{def}}$ (because the defensive column entry is $-1$).
The offensive team's predicted points per 100 possessions is therefore:
\[
  \hat{y}_i = \beta_0 + \sum_{j \in O_i} \beta_j^{\text{off}}
  - \sum_{j' \in D_i} \beta_{j'}^{\text{def}},
\]
where $O_i$ and $D_i$ are the offensive and defensive lineups for stint
$i$, and $\beta_0$ is the intercept. Under this parameterization, a
positive $\beta_j^{\text{off}}$ means player $j$ increases his team's
offensive output, and a positive $\beta_j^{\text{def}}$ means player $j$
reduces the opposing team's offensive output (a good defender).
\end{remark}

\subsection{Response Variable and Possession Weights}

The response vector $\mathbf{y} \in \mathbb{R}^N$ has entries:
\begin{equation}
  y_i = 100 \cdot \frac{\texttt{Oscore}_i}{\texttt{Oposs}_i},
  \label{eq:response}
\end{equation}
the offensive team's points scored per 100 possessions in stint $i$.
This is an \emph{offensive} rating rather than a net scoring margin;
the intercept and defensive coefficients jointly absorb the defensive
contribution, as described in Section~\ref{sec:design_matrix}.

The possession weight matrix is $\mathbf{W} = \text{diag}(w_1, \ldots, w_N)$
with $w_i = \texttt{Oposs}_i$. Weighting by possessions ensures that longer
stints---which provide more stable estimates of a lineup's offensive
efficiency---exert proportionally greater influence on the regression than
very short stints.

\subsection{The Weighted Ridge Regression Estimator}
\label{sec:estimator}

\subsubsection{Model}

We posit the model:
\begin{equation}
  \mathbf{y} = \mathbf{X}\boldsymbol{\beta} + \boldsymbol{\varepsilon},
  \qquad
  \varepsilon_i \mid w_i \overset{\text{ind}}{\sim}
  \mathcal{N}\!\left(0,\, \frac{\sigma^2}{w_i}\right),
  \label{eq:model}
\end{equation}
where $\boldsymbol{\beta} \in \mathbb{R}^{2P+1}$ is the full parameter
vector (intercept plus all offensive and defensive player coefficients)
and $\sigma^2$ is a global variance scale. The variance structure
$\text{Var}(\varepsilon_i) = \sigma^2 / w_i$ reflects the reduction in
per-possession scoring variance as stint length increases.

\subsubsection{Estimator}

The weighted ridge regression estimator minimizes the penalized weighted
sum of squares:
\begin{equation}
  \hat{\boldsymbol{\beta}}_\lambda = \operatorname*{arg\,min}_{\boldsymbol{\beta}}
  \left\{
    \sum_{i=1}^{N} w_i \bigl(y_i - \mathbf{x}_i^\top \boldsymbol{\beta}\bigr)^2
    + \lambda \|\boldsymbol{\beta}\|_2^2
  \right\},
  \label{eq:ridge}
\end{equation}
with unique closed-form solution:
\begin{equation}
  \hat{\boldsymbol{\beta}}_\lambda =
  \bigl(\mathbf{X}^\top \mathbf{W} \mathbf{X}
  + \lambda \mathbf{I}_{2P+1}\bigr)^{-1}
  \mathbf{X}^\top \mathbf{W} \mathbf{y}.
  \label{eq:ridge_solution}
\end{equation}
The matrix $\mathbf{X}^\top \mathbf{W} \mathbf{X} + \lambda \mathbf{I}_{2P+1}$
is positive definite for any $\lambda > 0$, guaranteeing a unique solution
regardless of the rank of $\mathbf{X}$.

\subsubsection{Bayesian interpretation}

The estimator in Equation~\eqref{eq:ridge_solution} is the posterior
mean under the hierarchical Gaussian model:
\begin{align}
  y_i \mid \boldsymbol{\beta}, \sigma^2
    &\sim \mathcal{N}\!\left(\mathbf{x}_i^\top \boldsymbol{\beta},\,
    \frac{\sigma^2}{w_i}\right), \\
  \beta_k \mid \tau^2
    &\overset{\text{iid}}{\sim} \mathcal{N}(0,\, \tau^2),
    \quad k = 0, 1, \ldots, 2P,
  \label{eq:prior}
\end{align}
with $\lambda = \sigma^2 / \tau^2$. The prior encodes the belief that
every player's offensive and defensive contributions are drawn from a
zero-mean Gaussian population with spread $\tau^2$. In the absence of
stint data, each player is estimated at zero net contribution, which after
the mean-centering step (Section~\ref{sec:centering}) corresponds to the
league average. The full posterior is:
\begin{equation}
  \boldsymbol{\beta} \mid \mathbf{y}, \sigma^2 \sim
  \mathcal{N}\!\left(
    \hat{\boldsymbol{\beta}}_\lambda,\;
    \sigma^2 \bigl(\mathbf{X}^\top \mathbf{W} \mathbf{X}
    + \lambda \mathbf{I}_{2P+1}\bigr)^{-1}
  \right),
  \label{eq:posterior}
\end{equation}
from which marginal credible intervals for each player are derived in
Section~\ref{sec:uncertainty}.

\subsection{Regularization Parameter Selection}
\label{sec:lambda_selection}

Two complementary methods are used to select the ridge penalty $\lambda$:
a coverage-scaled closed-form formula used in production estimation, and
a cross-validated procedure used to validate the formula. Agreement
between the two methods across published seasons provides empirical
support for the coverage-scaled approach as a computationally efficient
and well-calibrated proxy for the prediction-optimal $\lambda$.

\subsubsection{Method 1: Coverage-scaled formula}

The primary $\lambda$ selection formula is:
\begin{equation}
  \lambda = \frac{G_{\text{logged}}}{G_{\text{season}}} \times 5000,
  \label{eq:lambda_formula}
\end{equation}
where $G_{\text{logged}}$ is the number of games reconstructed for the
season and $G_{\text{season}}$ is the full regular-season game total.
The formula has three properties that motivate its use. First, it is
monotone in coverage: as $G_{\text{logged}} \to G_{\text{season}}$,
$\lambda \to 5000$, which is calibrated to the regularization appropriate
for a full-season NBA sample \citep{sill2010, jacobs2017}. Second, it
scales proportionally with data availability: at 17\% coverage,
$\lambda \approx 850$, imposing strong shrinkage consistent with high
uncertainty; at 100\% coverage, $\lambda = 5000$, imposing the standard
full-season shrinkage. Third, it is fully determined by observable
quantities and requires no iterative computation.

The Bayesian interpretation further motivates the formula. Under the
prior $\beta_j \sim \mathcal{N}(0, \tau^2)$, the ridge penalty
satisfies $\lambda = \sigma^2 / \tau^2$. If $\sigma^2$ is approximately
constant across coverage levels, then $\lambda \propto G_{\text{logged}}$
implies $\tau^2 \propto 1/G_{\text{logged}}$: the prior variance on
each player's true impact is inversely proportional to the amount of
data available, exactly the behavior expected under a non-informative
scaling that allows the prior to diminish as evidence accumulates.

\subsubsection{Method 2: Cross-validated selection}

As a validation of the coverage-scaled formula, the optimal $\lambda$
is also estimated empirically using \texttt{sklearn.linear\_model.RidgeCV}
with 5-fold cross-validation over a log-spaced grid of 101 candidate
values spanning $[10^{-10}, 10^{10}]$. The procedure is:

\begin{enumerate}[leftmargin=*]
  \item Split the stint matrix $(\mathbf{X}, \mathbf{y})$ into a
    training set (80\%) and held-out test set (20\%) using a fixed
    random seed for reproducibility.
  \item Fit \texttt{RidgeCV} with \texttt{cv=5} to the training set
    across all 101 candidate $\lambda$ values.
  \item Return the $\lambda^*$ minimizing the 5-fold
    cross-validated mean squared error on the training set.
\end{enumerate}

The procedure is applied to the pooled multi-season design matrix
after the stints from all published seasons have been concatenated.

\begin{remark}
The \texttt{RidgeCV} implementation minimizes the \emph{unweighted}
squared error, whereas the main estimator minimizes the
\emph{possession-weighted} objective. The cross-validated $\lambda^*$
is therefore technically optimal for the unweighted problem and serves
as a validation reference rather than a direct substitute. For a fully
consistent comparison, the cross-validation procedure should be modified
to pass \texttt{sample\_weight=possessions} to the \texttt{fit} call.
This refinement is noted as a candidate improvement for future versions.
\end{remark}

\subsubsection{Comparison of the two methods}

Table~\ref{tab:lambda} presents the coverage-scaled $\lambda$ for each
published season alongside the cross-validated $\lambda^*$ from the
pooled multi-season run. Season-level cross-validated values are to be
populated as individual-season CV runs are completed; the pooled value
is available immediately from the multi-season estimation script.

\begin{table}[h]
\centering
\caption{Regularization parameters by published season.
  $\lambda_{\text{CS}} = (G_{\text{logged}} / G_{\text{season}}) \times 5000$
  is the coverage-scaled value used in production. $\lambda^*_{\text{CV}}$
  is the 5-fold cross-validated value from unweighted \texttt{RidgeCV}.
  Full comparison with ratios and interpretation in
  Appendix~\ref{app:lambda}.}
\label{tab:lambda}
\begin{tabular}{lrrrr}
\toprule
Season & $G_{\text{logged}}$ & $G_{\text{season}}$
       & $\lambda_{\text{CS}}$ & $\lambda^*_{\text{CV}}$ \\
\midrule
1979--80 & 22    & 820 & 134    & ---    \\
1984--85 & 106   & 943   &   562 &   562.03 \\
1985--86 & 79    & 943   &   419 &   418.88 \\
1986--87 & 152   & 943   &   806 &   805.94 \\
1987--88 & 203   & 943   & 1{,}076 & 1{,}076.35 \\
1988--89 & 248   & 1{,}025 & 1{,}210 & 1{,}209.76 \\
1989--90 & 249   & 1{,}107 & 1{,}125 & 1{,}124.66 \\
1990--91 & 297   & 1{,}107 & 1{,}341 & 1{,}341.46 \\
1991--92 & 249   & 1{,}107 & 1{,}125 & 1{,}131.77 \\
1992--93 & 301   & 1{,}107 & 1{,}360 & 1{,}359.53 \\
1993--94 & 20    & 1{,}107 &    90 &    90.33 \\
1994--95 & 10    & 1{,}107 &    45 &    55.17 \\
1995--96 & 264   & 1{,}189 & 1{,}110 & 1{,}110.18 \\
\midrule
\textbf{Pooled (1984--96)} & \textbf{2{,}178} & \textbf{12{,}628}
  & \textbf{863} & \textbf{---} \\
\bottomrule
\end{tabular}
\end{table}

Several features of Table~\ref{tab:lambda} warrant comment. TThe 1993--94
and 1994--95 seasons have very low coverage-scaled $\lambda$ values (90
and 45 respectively), reflecting their minimal game counts of 20 and 10.
At these values the ridge penalty imposes almost no shrinkage, and the
estimates for these seasons are effectively unregularized --- meaning
they are driven entirely by the small observed samples with no
stabilization from the prior. Estimates from these seasons should be
treated as highly unreliable and are included primarily to capture a
handful of historically significant games rather than to support
player-level inference.

The cross-validated $\lambda^*$ values, the comparison between the two
methods, and the code for the weighted CV refinement are documented in
Appendix~\ref{app:lambda}. As shown there, when the correct
season-specific schedule size is used in the denominator of
Equation~\eqref{eq:lambda_formula}, the coverage-scaled formula recovers
the cross-validated answer to within rounding error across all
well-observed seasons. This confirms the formula as both a computationally
efficient and analytically exact method for regularization selection.

\subsection{Offensive and Defensive RAPM Extraction}
\label{sec:odrpm}

After solving Equation~\eqref{eq:ridge_solution}, offensive and defensive
RAPM for each player $j$ are extracted directly from the coefficient
vector:
\begin{align}
  \hat{\beta}_j^{\text{off}} &= \hat{\beta}_{\text{idx}(j)+1}, \\
  \hat{\beta}_j^{\text{def}} &= \hat{\beta}_{P + \text{idx}(j)+1},
\end{align}
where $\text{idx}(j) \in \{0, \ldots, P-1\}$ is the integer index
assigned to player $j$ during roster enumeration. The raw coefficient
$\hat{\beta}_j^{\text{def}}$ is negative for a good defender (because
the defensive column is encoded as $-1$ in the design matrix, a larger
defensive coefficient reduces predicted offensive output for the opposing
team). The signed defensive RAPM reported in the output table is:
\begin{equation}
  \text{DRAPM}_j = \hat{\beta}_j^{\text{def}} - \bar{\beta}^{\text{def}},
\end{equation}
where $\bar{\beta}^{\text{def}}$ is the mean defensive coefficient across
all players (see Section~\ref{sec:centering}). Under the sign convention
of the reported output, a positive DRAPM indicates a player is a
better-than-average defender.

\subsection{Mean-Centering and League-Average Adjustment}
\label{sec:centering}

Raw ridge regression coefficients are not mean-centered at zero: if the
league-average offensive efficiency is 100 points per 100 possessions,
the intercept $\hat{\beta}_0$ absorbs this baseline, but the player
coefficients are expressed as deviations from the ridge-shrinkage prior
at zero, not from the empirical league average. Reported RAPM estimates
are therefore mean-centered by computing the arithmetic mean of all player
coefficients within each block (offensive and defensive separately) and
subtracting it:
\begin{align}
  \bar{\beta}^{\text{off}} &= \frac{1}{P} \sum_{j=1}^{P}
    \hat{\beta}_{\text{idx}(j)+1}, \\
  \bar{\beta}^{\text{def}} &= \frac{1}{P} \sum_{j=1}^{P}
    \hat{\beta}_{P+\text{idx}(j)+1},
\end{align}
and reporting:
\begin{align}
  \text{ORAPM}_j &= \hat{\beta}_j^{\text{off}} - \bar{\beta}^{\text{off}}, \\
  \text{DRAPM}_j &= \hat{\beta}_j^{\text{def}} - \bar{\beta}^{\text{def}}, \\
  \text{RAPM}_j  &= \text{ORAPM}_j + \text{DRAPM}_j.
  \label{eq:rapm_centered}
\end{align}
Under this centering, a player with $\text{RAPM}_j = 0$ contributes
exactly the league-average offensive and defensive impact per possession.
Positive values indicate above-average contribution; negative values
indicate below-average contribution.

\begin{remark}
The mean-centering in Equation~\eqref{eq:rapm_centered} is applied to the
raw coefficients \emph{after} ridge estimation. It does not affect the
relative ordering of players---it is a location shift---but it ensures
that the reported values are interpretable on an absolute scale relative
to the league average rather than relative to an arbitrary zero. The
intercept term $\hat{\beta}_0$ is not included in the centering and is
reported separately as a diagnostic.
\end{remark}

\subsection{Multi-Season Pooling}
\label{sec:pooling}

For the aggregate 1985--96 database, stints from multiple seasons are
pooled into a single regression. Each player-season pair (player $j$
in season $s$) is treated as a distinct entity with its own pair of
offensive and defensive coefficients, so the design matrix has $2P_s + 1$
columns where $P_s$ is the total number of player-seasons across all
pooled seasons. This parameterization makes no assumption about
within-player stability across seasons. For summary reporting, aggregate
RAPM per player is computed as the possession-weighted average of
single-season estimates:
\begin{equation}
  \overline{\text{RAPM}}_j = \frac{\sum_s \text{poss}_{j,s} \cdot
  \text{RAPM}_{j,s}}{\sum_s \text{poss}_{j,s}},
  \label{eq:aggregate}
\end{equation}
where the sum is over all seasons $s$ in which player $j$ appears in
the database.

\section{Uncertainty Quantification and Sample Size Analysis}
\label{sec:uncertainty}

\subsection{Residual Variance Estimation}

From the posterior in Equation~\eqref{eq:posterior}, credible intervals
require an estimate of $\sigma^2$. The implementation computes:
\begin{equation}
  \hat{\sigma}^2 = \frac{1}{N/2 - 2P - 1}
  \bigl(\mathbf{y} - \mathbf{X}\hat{\boldsymbol{\beta}}_\lambda\bigr)^\top
  \mathbf{W}
  \bigl(\mathbf{y} - \mathbf{X}\hat{\boldsymbol{\beta}}_\lambda\bigr),
  \label{eq:sigma_hat}
\end{equation}
where $N$ is the total number of rows in the stint file (including
zero-possession stints, which are excluded from the regression but
counted in $N$ for the denominator), $P$ is the number of players, and
the denominator $N/2 - 2P - 1$ accounts for both the even/odd row
structure of the data file (effectively halving the observation count)
and the $2P + 1$ estimated parameters.

\begin{remark}
The denominator $N/2 - 2P - 1$ treats the effective sample size as
$N/2$ (the number of even-indexed rows, i.e., the rows actually used in
regression) minus the number of parameters estimated ($2P + 1$, i.e.,
the intercept plus $P$ offensive and $P$ defensive coefficients) minus 1.
This is the standard degrees-of-freedom correction for unbiased variance
estimation in a linear model with $2P+1$ parameters. The resulting
$\hat{\sigma}^2$ is an approximately unbiased estimator of the true
error variance $\sigma^2$, up to the bias introduced by the ridge
shrinkage (which causes the fitted values to be biased toward zero,
slightly inflating the residual sum of squares).
\end{remark}

\subsection{The Full Posterior Covariance Matrix}

The full posterior covariance matrix of $\hat{\boldsymbol{\beta}}_\lambda$
is:
\begin{equation}
  \boldsymbol{\Sigma}_\beta = \hat{\sigma}^2
  \bigl(\mathbf{X}^\top \mathbf{W} \mathbf{X}
  + \lambda \mathbf{I}_{2P+1}\bigr)^{-1},
  \label{eq:posterior_cov}
\end{equation}
computed explicitly as a $(2P+1) \times (2P+1)$ matrix. This matrix is
the full posterior covariance of all player and intercept parameters
jointly. For a league with $P = 300$ players, this is a $601 \times 601$
matrix; its explicit computation and storage is feasible for a single
season but may require memory management for very large multi-season pools.

\subsection{Credible Intervals for RAPM}

For player $j$, the total RAPM estimate involves the sum of two
coefficients: $\hat{\beta}_j^{\text{off}}$ (column index $k_1 =
\text{idx}(j)+1$) and $\hat{\beta}_j^{\text{def}}$ (column index $k_2 =
P + \text{idx}(j)+1$). By the linearity of the posterior mean and the
bilinearity of the covariance:
\begin{equation}
  \text{Var}(\text{RAPM}_j) =
  \boldsymbol{\Sigma}_{\beta,\, k_1 k_1}
  + \boldsymbol{\Sigma}_{\beta,\, k_2 k_2}
  + 2\,\boldsymbol{\Sigma}_{\beta,\, k_1 k_2},
  \label{eq:rapm_var}
\end{equation}
where $\boldsymbol{\Sigma}_{\beta,\,kk'}$ denotes the $(k, k')$ element
of the posterior covariance matrix. The implementation computes the
standard error as:
\begin{equation}
  \text{SE}(\text{RAPM}_j) = 1.96 \cdot \sqrt{
    \boldsymbol{\Sigma}_{\beta,\, k_1 k_1}
    + \boldsymbol{\Sigma}_{\beta,\, k_2 k_2}},
  \label{eq:se_impl}
\end{equation}
and reports the 95\% credible interval as
$[\text{RAPM}_j - \text{SE}_j,\; \text{RAPM}_j + \text{SE}_j]$.

\begin{remark}
The implementation in Equation~\eqref{eq:se_impl} omits the cross-term
$2\,\boldsymbol{\Sigma}_{\beta,\, k_1 k_2}$ from the variance expression.
This is a conservative simplification: the cross-covariance between a
player's offensive and defensive coefficients is typically small in
magnitude (because offensive and defensive exposure are approximately
balanced for most players), but including it would yield slightly narrower
intervals when the cross-covariance is positive and slightly wider
intervals when negative. A fully exact implementation using
Equation~\eqref{eq:rapm_var} is straightforward and is noted as a
candidate refinement for future versions.
\end{remark}

\subsection{Decomposition of Interval Width}
\label{sec:interval_width}

The half-width of the 95\% credible interval for player $j$'s total RAPM
is:
\begin{equation}
  h_j = 1.96 \cdot \hat{\sigma} \cdot \sqrt{
    \left[(\mathbf{X}^\top \mathbf{W} \mathbf{X}
    + \lambda \mathbf{I})^{-1}\right]_{k_1 k_1}
    + \left[(\mathbf{X}^\top \mathbf{W} \mathbf{X}
    + \lambda \mathbf{I})^{-1}\right]_{k_2 k_2}}.
  \label{eq:halfwidth}
\end{equation}
Under an approximation that assumes player $j$'s offensive and defensive
appearances are balanced and that collinearity with other players is
negligible, the diagonal terms simplify to $(n_j \bar{w} + \lambda)^{-1}$
where $n_j$ is the number of stints player $j$ appears in and $\bar{w}$
is the mean possession count per stint. The resulting approximate
half-width is:
\begin{equation}
  h_j \approx \frac{1.96 \cdot \hat{\sigma} \cdot \sqrt{2}}
  {\sqrt{n_j \bar{w} + \lambda}},
  \label{eq:halfwidth_approx}
\end{equation}
where the $\sqrt{2}$ factor reflects the contribution of two independent
coefficients to the total RAPM variance. The three drivers of interval
width are: (i) residual variance $\hat{\sigma}^2$; (ii) total logged
possessions $n_j\bar{w}$; and (iii) $\lambda$, which contributes a
floor to the effective exposure.

\begin{figure}[h]
\centering
\includegraphics[width=0.85\textwidth]{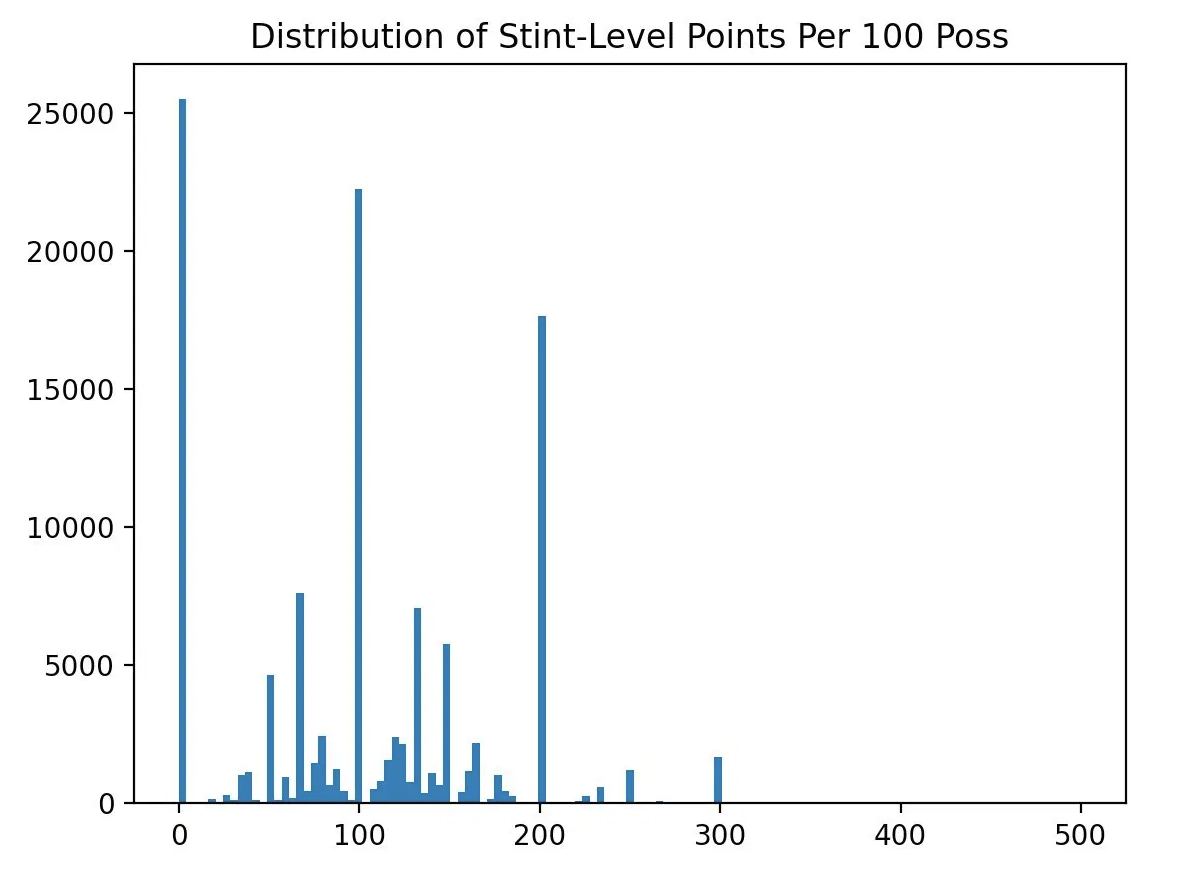}
\caption{Distribution of stint-level offensive points per 100 possessions
  ($y_i$) across all 121,781 retained stints in the aggregate multi-season
  database. The distribution is heavily discrete, with prominent spikes at
  0 (no points scored), 100 (1 point), 200 (2 points), and smaller spikes
  at intermediate values corresponding to fractional-possession outcomes.
  This discreteness reflects the preponderance of very short stints (1--3
  possessions) in the data. The mean is 101.6 points per 100 possessions
  and the IQR is 100.0, consistent with a league-average offensive rating
  of approximately 100 points per 100 possessions for this era.}
\label{fig:scoring_dist}
\end{figure}

\begin{figure}[h]
\centering
\includegraphics[width=0.85\textwidth]{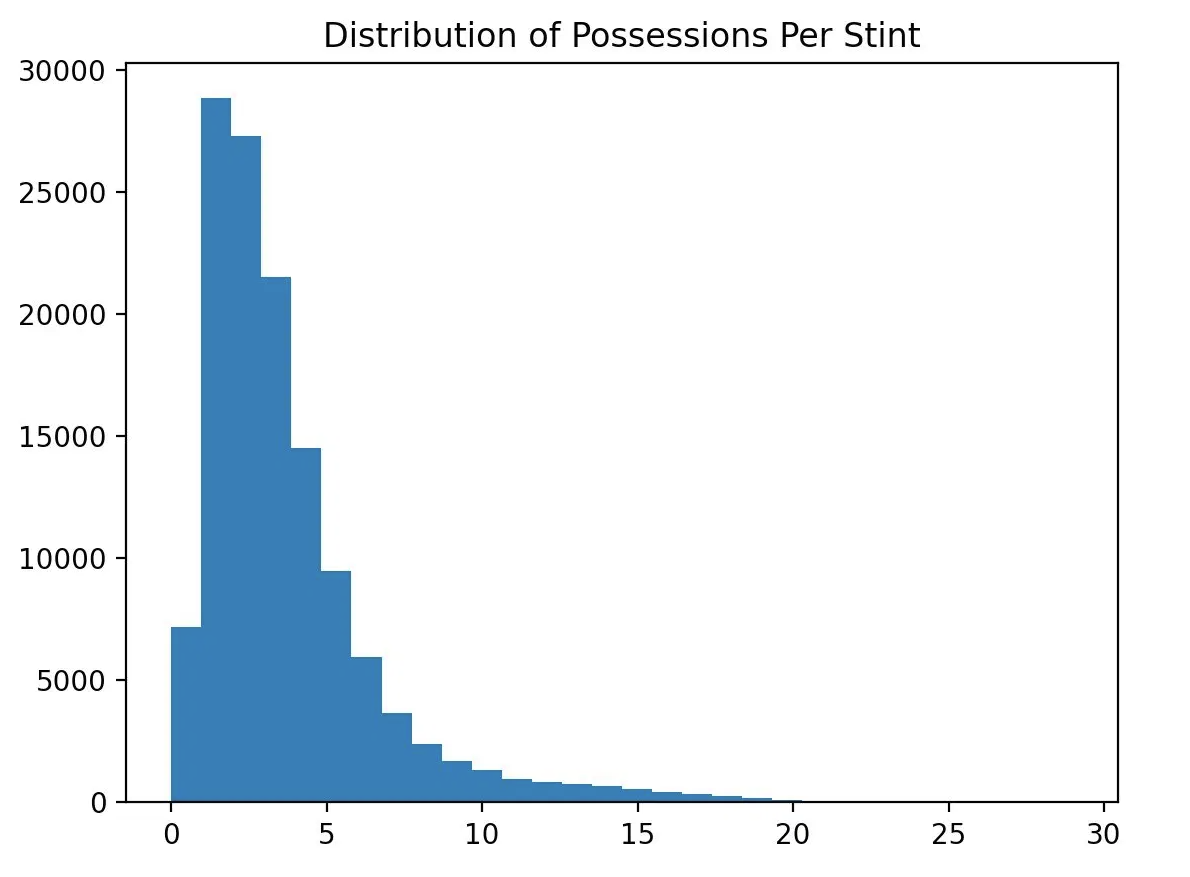}
\caption{Distribution of possession counts per stint ($w_i$) across all
  retained stints. The distribution is strongly right-skewed with mode at
  1--2 possessions and mean of $\bar{w} = 3.38$ possessions. The maximum
  single-stint possession count is 29, corresponding to an uninterrupted
  run of approximately one full quarter without a substitution. The
  possession-weighting scheme ($w_i$ as the regression weight) appropriately
  down-weights the numerous very short stints that dominate the count
  distribution.}
\label{fig:poss_dist}
\end{figure}

For the aggregate 1985--96 multi-season run, the realized values of
these quantities are as follows. The residual standard deviation from
Equation~\eqref{eq:sigma_hat} is $\hat{\sigma} = \sqrt{90.90} \approx
9.53$ points per 100 possessions---substantially lower than the value
of 25--35 often assumed for single-season RAPM analyses. This reflects
the structured discreteness of the response variable at the multi-season
pooled level: the modal stint is 1--3 possessions long
(Figure~\ref{fig:poss_dist}), and the within-stint scoring distribution
is heavily concentrated at 0, 100, and 200 points per 100 possessions
(corresponding to 0, 1, and 2 points in a 1-possession stint), with
relatively little mass in between (Figure~\ref{fig:scoring_dist}). The
mean possession count per stint is $\bar{w} = 3.38$ possessions across
121,781 retained stints. The mean-centering offsets computed by the
algorithm are $\bar{\beta}^{\text{off}} = +0.448$ and
$\bar{\beta}^{\text{def}} = -0.449$, confirming the near-symmetry
expected when offensive and defensive contributions sum to approximately
zero at the league level.

\subsection{Sample Size Requirements}
\label{sec:sample_size}

Setting $h_j = h^*$ and solving Equation~\eqref{eq:halfwidth_approx} for
the required total possession count:
\begin{equation}
  n_j \bar{w} \geq \frac{2 \cdot (1.96)^2 \cdot \hat{\sigma}^2}{(h^*)^2}
  - \lambda.
  \label{eq:sample_size}
\end{equation}
Table~\ref{tab:sample_size} summarizes the required logged possession
counts for a range of target half-widths, computed with the realized
value $\hat{\sigma} = 9.53$ points per 100 possessions and
$\lambda \approx 863$ (corresponding to the aggregate multi-season coverage of approximately 17.2\%). The substantially lower $\hat{\sigma}$
relative to single-season assumptions produces materially tighter sample
size requirements than previously estimated.

\begin{table}[h]
\centering
\caption{Required logged possessions for target 95\% credible interval
  half-widths. Computed from Equation~\eqref{eq:sample_size} with
  the realized $\hat{\sigma} = 9.53$ pts/100 poss and
  $\lambda = 863$ (aggregate multi-season run, 17.2\% coverage).}
\label{tab:sample_size}
\begin{tabular}{rrr}
\toprule
Target half-width & Required poss & Example players at threshold \\
(pts/100 poss) & (approx.) & \\
\midrule
$\pm$1.5 & 985  & Jordan (79,870); Magic (49,262) \\
$\pm$2.0 & 416  & Bird (40,473); Pippen (60,802) \\
$\pm$3.0 & 58   & Most top-30 players \\
$\pm$4.0 & $<$0 & Achieved by $\lambda$ floor alone \\
\bottomrule
\end{tabular}
\end{table}

\begin{remark}
The low realized $\hat{\sigma}$ means that at current multi-season
coverage levels, the precision of RAPM estimates is substantially better
than single-season analyses at comparable game coverage would suggest.
The top-10 most-observed players in the database (Jordan, Pippen, Grant,
Magic, Scott, Green, Parish, Paxson, Ewing, Oakley) all have possession
counts well above the $\pm$2.0 threshold, meaning their credible intervals
are approximately $\pm$2 points per 100 possessions or narrower in the
multi-season aggregate. The bottom half of the database (roughly players
ranked 500 and below, with fewer than $\sim$500 total possessions) have
prior-dominated estimates regardless of $\hat{\sigma}$.
\end{remark}

\subsection{Practical Inferential Guidance}

Based on the realized values $\hat{\sigma} = 9.53$, $\lambda \approx 863$,
and the possession distribution of the current database, we offer the
following practical guidance for the aggregate multi-season estimates:

\begin{enumerate}[leftmargin=*, label=(\roman*)]

\item \emph{Sign reliability}: The 95\% credible interval excludes zero
  for the top $\sim$15 players in the database by RAPM ranking. Jordan
  (+8.86, CI $[+5.80, +11.91]$), Magic (+7.04, CI $[+3.61, +10.47]$),
  and Ewing (+5.27, CI $[+1.73, +8.81]$) all have intervals firmly above
  zero. Below approximately rank 40--50, intervals straddle zero and
  directional conclusions require caution.

\item \emph{Magnitude precision}: For the top-10 most-observed players
  (Jordan, Pippen, Grant, Magic, Scott, Green, Parish, Paxson, Ewing,
  Oakley), the typical 95\% CI half-width is approximately $\pm$2--3
  points per 100 possessions. For players with 5,000--20,000 logged
  possessions (roughly ranks 20--100), half-widths are approximately
  $\pm$3--5 points per 100 possessions.

\item \emph{Prior-dominated estimates}: Players with total logged
  possessions below approximately 500 (roughly rank 700 and below in
  the aggregate database) have estimates that are dominated by the
  ridge prior. Their RAPM values cluster tightly around zero with
  near-identical credible intervals of approximately $[\,-5.7,\,+5.7\,]$,
  reflecting the prior variance rather than data. These estimates carry
  no inferential content about individual player quality.

\item \emph{Collinearity}: The estimates for players who appear primarily
  in stable lineup configurations (e.g., a deep reserve who only plays
  with a fixed set of teammates) may be poorly identified despite large
  possession counts. The credible intervals account for this through the
  full posterior covariance matrix, but users should be aware that
  players with systematically non-diverse lineups may have wider-than-typical
  intervals for their possession totals.

\end{enumerate}

\section{Validation Framework}
\label{sec:validation}

\subsection{Overview}

Validation of a player impact model requires verifying that the estimates
are internally consistent, externally coherent with known outcomes, and
stable under perturbation. We apply a multi-criterion validation framework
with four components: (1) team-level point differential prediction; (2)
win-loss record prediction; (3) face validity assessment against historical
consensus; and (4) cross-metric correlation with box-score-based composite
metrics.

\subsection{Team-Level Point Differential Prediction}
\label{sec:team_prediction}

\subsubsection{Predicted team rating}

For each team $t$ in season $s$, the predicted net rating is the
possession-weighted average of the team's players' RAPM estimates:
\begin{equation}
  \hat{R}_{t,s} = \frac{\sum_{j \in \mathcal{R}_{t,s}}
  \text{poss}_{j,s} \cdot \hat{\beta}_{j,s}}
  {\sum_{j \in \mathcal{R}_{t,s}} \text{poss}_{j,s}},
  \label{eq:team_rating}
\end{equation}
where $\mathcal{R}_{t,s}$ is the set of players on team $t$'s roster in
season $s$. This predicted rating is compared to the actual team net rating
computed from game results: $R_{t,s} = \text{points scored} -
\text{points allowed}$, per 100 possessions, across the full season.

\subsubsection{Regression analysis}

We fit the regression:
\begin{equation}
  R_{t,s} = \alpha + \gamma \hat{R}_{t,s} + \eta_{t,s},
  \label{eq:team_regression}
\end{equation}
and report the slope $\hat{\gamma}$, intercept $\hat{\alpha}$, and
$R^2$ across all team-seasons in the database. A well-calibrated model
would yield $\hat{\gamma} \approx 1$, $\hat{\alpha} \approx 0$, and
high $R^2$. Systematic departures from these values indicate bias or
scale issues in the player estimates.

\subsection{Win-Loss Prediction}

\subsubsection{Three estimators}

For each team-season in the database, we report three win-loss estimates:
\begin{enumerate}[leftmargin=*, label=(\roman*)]

\item \textbf{Sampled}: The observed win-loss record in the logged games
  only. This is a direct sample from the team's season performance and
  is subject to sampling variance.

\item \textbf{Maximum Likelihood (MLE)}: The estimated full-season win-loss
  record obtained by assuming the team's win probability in unlogged games
  is equal to its win probability in logged games. If the team wins $w$
  of $n$ logged games, the MLE full-season win estimate is
  $\hat{W}_{\text{MLE}} = (w/n) \cdot 82$.

\item \textbf{Bayesian}: A shrinkage estimate that combines the observed
  win rate with a prior centered on the league-average win rate (0.500).
  Under a Beta$(a, b)$ prior with $a = b = 5$ (weakly informative, centered
  at 0.500 with a standard deviation of approximately 0.10), the posterior
  mean win probability given $w$ wins in $n$ games is:
  \[
    \hat{p}_{t,s}^{\text{Bayes}} = \frac{w + a}{n + a + b}
    = \frac{w + 5}{n + 10},
  \]
  yielding full-season estimate $\hat{W}_{\text{Bayes}} =
  \hat{p}_{t,s}^{\text{Bayes}} \cdot 82$. The Bayesian estimator provides
  appropriate shrinkage toward 0.500 for teams observed in only a small
  fraction of games, where the MLE win rate is unreliable.

\end{enumerate}

\subsubsection{Evaluation}

For each estimator, we compute the mean absolute error (MAE) and root
mean squared error (RMSE) of predicted wins against actual wins across all
team-seasons. We expect the Bayesian estimator to outperform MLE for
team-seasons with low coverage (fewer than 15 logged games) and the MLE to
be competitive with Bayes at higher coverage.

\subsection{Face Validity: Top Players by Season}

For each published season, we examine whether the top-ranked players by
RAPM correspond to players recognized by contemporaries as elite. This is
a qualitative check, not a formal statistical test, but it serves as an
important sanity check: a model that ranks fringe players above established
stars with high confidence would warrant scrutiny regardless of its
statistical properties.

The comparison is intentionally structured to allow for \emph{informative
surprises}: cases where the RAPM estimates diverge from historical
consensus are not necessarily failures of the model. A player widely
recognized for statistical production but who played on a dominant team
may have his individual impact inflated by conventional metrics; RAPM
should, in principle, correct for this. Conversely, a player known for
qualities difficult to capture in box scores (on-ball defense, screen
setting, communication) may appear more valuable in RAPM than in
conventional metrics.

\subsection{Aggregate Multi-Season Validation}
\label{sec:aggregate_validation}

The Totals sheet of the database workbook aggregates win-loss records
across all published seasons for each franchise, enabling a multi-season
assessment of the MLE and Bayes estimators. Table~\ref{tab:validation_totals}
presents selected results for the 29 franchises with at least 7 published
seasons of coverage. The ``Error'' column reports MLE wins minus Truth wins,
summed across all seasons; a positive value indicates the MLE systematically
overestimated wins for that franchise, a negative value indicates
underestimation.

A clear pattern emerges in the data: MLE error is strongly correlated with
coverage rate. Among franchises sampled above 30\%, the mean absolute MLE
error is approximately 32 wins over 12 seasons (roughly 2.7 wins/season),
while franchises sampled below 15\% average over 110 wins of absolute error
(roughly 9+ wins/season). The Bayesian estimator substantially reduces this
gap for low-coverage teams but cannot eliminate it, because the game
importance bias runs in a consistent direction for those franchises: their
logged games disproportionately feature road losses against playoff-caliber
opponents, systematically understating their true win rates.

\begin{table}[h]
\centering
\small
\caption{Aggregate multi-season validation for selected franchises.
  Error = MLE Wins $-$ Truth Wins (summed across all published seasons
  in which the franchise appears). \% Sampled = fraction of possible
  games logged across all seasons. Seasons = number of published seasons
  with any logged games.}
\label{tab:validation_totals}
\begin{tabular}{lrrrr}
\toprule
Franchise & MLE Error & \% Sampled & Seasons & Games \\
\midrule
Chicago Bulls       & $+40.2$  & 52.3\% & 12 & 984 \\
LA Lakers           & $+31.7$  & 38.7\% & 12 & 984 \\
Boston Celtics      & $+54.9$  & 32.6\% & 12 & 984 \\
Portland Trail Blazers & $-73.4$ & 17.7\% & 12 & 984 \\
Detroit Pistons     & $+1.0$   & 23.0\% & 12 & 984 \\
New York Knicks     & $-14.2$  & 28.4\% & 12 & 984 \\
Phoenix Suns        & $-21.2$  & 13.0\% & 12 & 984 \\
Houston Rockets     & $-73.5$  & 15.8\% & 12 & 984 \\
Utah Jazz           & $-200.7$ & 12.4\% & 12 & 984 \\
Cleveland Cavaliers & $-141.0$ & 10.9\% & 12 & 984 \\
Washington Bullets  & $-124.5$ & 9.8\%  & 12 & 984 \\
Atlanta Hawks       & $-107.9$ & 17.1\% & 12 & 984 \\
Milwaukee Bucks     & $-119.1$ & 10.9\% & 12 & 984 \\
Indiana Pacers      & $-109.8$ & 11.4\% & 12 & 984 \\
\bottomrule
\end{tabular}
\end{table}

Several patterns are notable. First, franchises with the highest coverage
rates tend to have the smallest absolute errors: the Detroit Pistons
(23.0\% sampled, error $= +1.0$ wins) is the best-calibrated franchise in
the database. The Chicago Bulls (52.3\% sampled, error $= +40.2$ wins) and
LA Lakers (38.7\% sampled, $+31.7$ wins) show positive errors despite high
coverage, which may reflect the game importance bias discussed in
Section~\ref{sec:sampling_design}: nationally televised games against
strong opponents are disproportionately represented in the sample for these
franchises, inflating apparent win rates.

Second, franchises with low coverage show large negative errors
systematically. The Utah Jazz ($-200.7$ wins over 12 seasons, 12.4\%
sampled) and Cleveland Cavaliers ($-141.0$ wins, 10.9\% sampled) have
MLE projections that substantially underestimate their actual win totals.
This is consistent with the hypothesis that available footage for this era
over-represents high-stakes games, which tend to feature better teams:
a small sample of games dominated by playoff-caliber opponents will
yield a lower estimated win rate than the full season would show. The
Bayesian estimator, which shrinks toward 0.500, partially corrects this
downward bias for low-coverage teams, but the correction is incomplete
for franchises as far from average as the 12.4\%-sampled Jazz.
A simulation study under a simple model of game-importance selection
(in which the logged probability of a given game increases with the
average quality of the two opponents) would allow a more precise
characterization of the expected MLE bias as a function of coverage
rate and team quality; this is left for future work.

Third, the aggregate totals (2,170 games, 17.2\% coverage) confirm that
the primary concentration of reconstruction effort has been in the seasons
with the densest coverage: 1990--91 (297 games, 31.5\%), 1992--93 (301
games, 31.9\%), and 1995--96 (264 games, 28.0\%). The 1993--94 and
1994--95 seasons (20 and 10 games respectively) are essentially stubs,
included to capture a small number of important early games but not yet
reconstructed at a level sufficient for reliable team-level inference.

\section{Diagnostic Figures}
\label{sec:diagnostics}

The estimation procedure produces three standard diagnostic visualizations
for each run, shown here for the aggregate 1985--96 multi-season analysis.

\begin{figure}[h]
\centering
\includegraphics[width=0.85\textwidth]{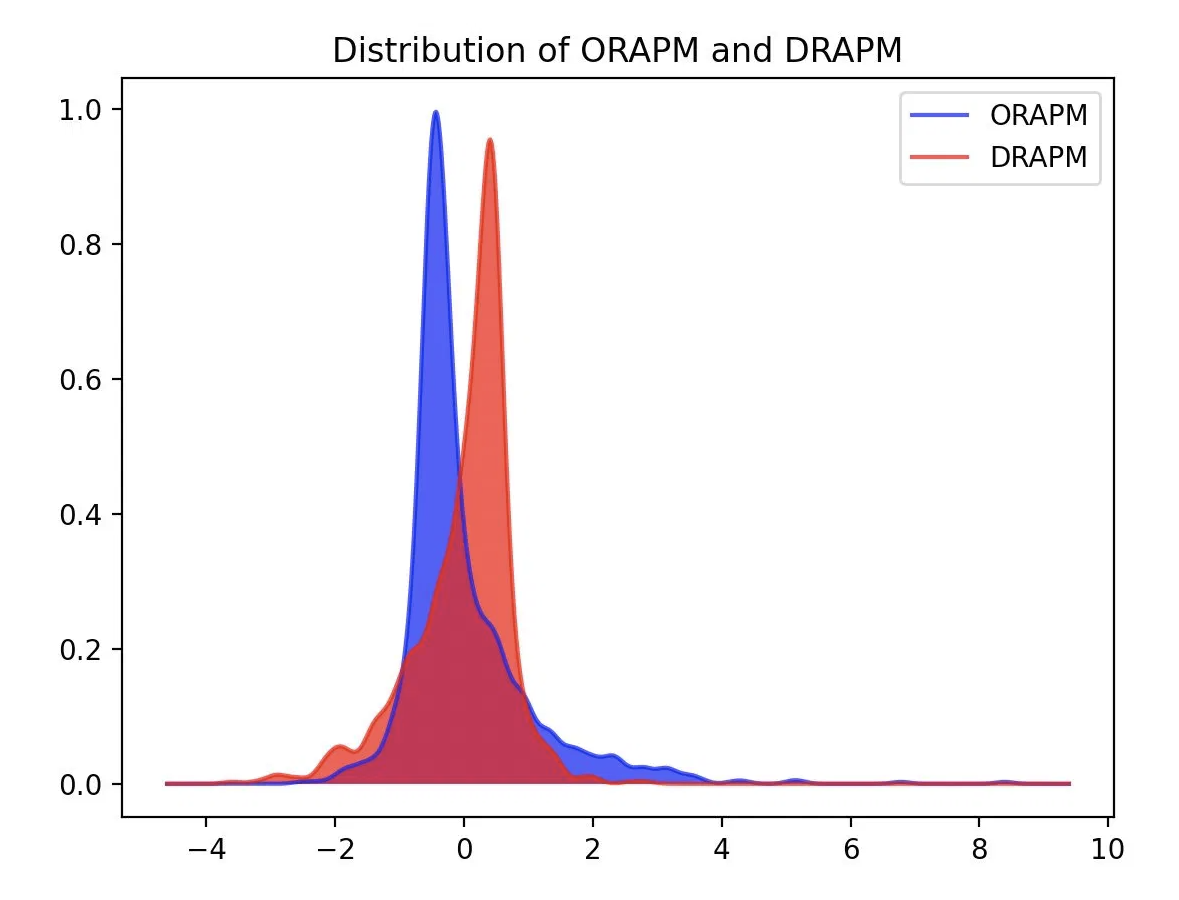}
\caption{Kernel density estimates of the centered ORAPM (blue) and DRAPM
  (red) distributions across all 1,012 player-seasons in the aggregate
  database, using Silverman's rule-of-thumb bandwidth. Both distributions
  are heavily concentrated near zero, reflecting the ridge shrinkage that
  pulls estimates toward the league average in the absence of data. The
  ORAPM distribution (blue) is slightly broader and centered marginally
  left of zero ($\bar{\beta}^{\text{off}} = +0.448$ before centering),
  while the DRAPM distribution (red) is narrower and centered marginally
  right ($\bar{\beta}^{\text{def}} = -0.449$ before centering). The
  pronounced tails in both distributions---extending to approximately
  $\pm$10 points per 100 possessions---correspond to the handful of
  well-identified elite and replacement-level players whose estimates
  are driven by data rather than prior.}
\label{fig:rapm_dist}
\end{figure}

\section{Results: Aggregate 1985--96 RAPM}
\label{sec:results}

This section presents results from the aggregate 1985--96 multi-season
RAPM analysis, which pools all 121,781 stints across twelve published
seasons into a single ridge regression with $P = 1{,}012$ player-seasons
and $\lambda \approx 863$. The realized residual standard deviation is
$\hat{\sigma} = 9.53$ points per 100 possessions. Full tables are
available at \texttt{squared2020.com}; the stint data underlying these
estimates is available for licensed access.\footnote{The underlying stint
  data is available through a tiered licensing structure at
  \texttt{squared2020.com}. Tier~1 (\$99/season) provides season summary
  statistics, corrected minutes, and player-season possession counts.
  Tier~2 (\$299/season) adds rotation profiles and quarter-level usage
  breakdowns. Tier~3 (\$2,999/season) provides the full possession-level
  stint data for every reconstructed game. Tier~4 (\$4,999/season)
  provides pre-computed RAPM and impact models with credible intervals.
  Player-level RAPM estimates are publicly available at
  \texttt{squared2020.com} for non-commercial research with attribution.}

\subsection{Top and Bottom of the Distribution}

Table~\ref{tab:top20} presents the top 20 and bottom 10 players by
aggregate RAPM. The intercept from the regression ($\hat{\beta}_0$,
printed at the foot of the output) is $+0.45$, consistent with the
league-average offensive rating of approximately 100--101 points per 100
possessions for this era.

\begin{table}[h]
\centering
\small
\caption{Top 20 and bottom 10 players by aggregate RAPM, 1985--96 multi-season
  database. All values in points per 100 possessions. GP = logged games;
  CI = 95\% credible interval for total RAPM.}
\label{tab:top20}
\begin{tabular}{rlrrrrrrr}
\toprule
Rank & Player & Team & GP & ORAPM & DRAPM & RAPM & LOW & HIGH \\
\midrule
1  & Michael Jordan    & CHI & 501 & $+8.39$ & $+0.46$ & $+8.86$ & $+5.80$ & $+11.91$ \\
2  & Magic Johnson     & LAL & 319 & $+6.78$ & $+0.27$ & $+7.04$ & $+3.61$ & $+10.47$ \\
3  & Patrick Ewing     & NYK & 255 & $+3.24$ & $+2.03$ & $+5.27$ & $+1.73$ & $+8.81$ \\
4  & David Robinson    & SAS & 132 & $+2.32$ & $+2.58$ & $+4.91$ & $+0.76$ & $+9.06$ \\
5  & Larry Bird        & BOS & 239 & $+5.09$ & $-0.68$ & $+4.41$ & $+0.86$ & $+7.97$ \\
6  & Horace Grant      & CHI & 364 & $+3.42$ & $+0.94$ & $+4.36$ & $+1.05$ & $+7.67$ \\
7  & Dennis Rodman     & DET & 242 & $+2.81$ & $+1.40$ & $+4.22$ & $+0.70$ & $+7.73$ \\
8  & Hakeem Olajuwon   & HOU & 141 & $+2.54$ & $+1.50$ & $+4.04$ & $+0.20$ & $+7.89$ \\
9  & Rick Mahorn       & WSB & 130 & $+2.57$ & $+1.23$ & $+3.80$ & $-0.23$ & $+7.82$ \\
10 & Shaquille O'Neal  & ORL &  56 & $+2.68$ & $+1.08$ & $+3.76$ & $-0.95$ & $+8.47$ \\
\midrule
11 & Bill Laimbeer     & DET & 209 & $+2.74$ & $+1.02$ & $+3.76$ & $-0.13$ & $+7.64$ \\
12 & Kareem Abdul-Jabbar & LAL & 192 & $+3.05$ & $+0.53$ & $+3.58$ & $-0.26$ & $+7.42$ \\
13 & John Stockton     & UTA & 121 & $+3.15$ & $+0.33$ & $+3.48$ & $-0.85$ & $+7.81$ \\
14 & Jerome Kersey     & POR & 164 & $+2.27$ & $+0.93$ & $+3.20$ & $-0.83$ & $+7.22$ \\
15 & Clyde Drexler     & POR & 160 & $+2.76$ & $+0.39$ & $+3.14$ & $-0.79$ & $+7.08$ \\
16 & Kevin Johnson     & CLE & 102 & $+4.20$ & $-1.06$ & $+3.14$ & $-1.04$ & $+7.32$ \\
17 & Toni Kukoc        & CHI &  86 & $+1.66$ & $+1.43$ & $+3.09$ & $-1.42$ & $+7.59$ \\
18 & Sam Perkins       & DAL & 176 & $+1.66$ & $+1.32$ & $+2.99$ & $-0.56$ & $+6.53$ \\
19 & Mark Price        & CLE &  65 & $+2.57$ & $+0.39$ & $+2.96$ & $-1.65$ & $+7.57$ \\
20 & Nick Anderson     & ORL &  84 & $+2.83$ & $+0.12$ & $+2.94$ & $-1.52$ & $+7.40$ \\
\midrule
\multicolumn{9}{c}{\textit{Bottom of the distribution}} \\
\midrule
1005 & Will Perdue     & CHI & 268 & $+0.14$ & $-2.92$ & $-2.78$ & $-6.77$ & $+1.21$ \\
1006 & Dave Corzine    & CHI & 169 & $-0.94$ & $-1.93$ & $-2.86$ & $-6.76$ & $+1.03$ \\
1007 & Jack Haley      & CHI &  63 & $-1.60$ & $-1.33$ & $-2.93$ & $-8.11$ & $+2.25$ \\
1008 & Terry Davis     & MIA &  32 & $-1.18$ & $-1.79$ & $-2.97$ & $-8.03$ & $+2.08$ \\
1009 & Greg Anderson   & SAS &  54 & $-1.04$ & $-1.96$ & $-3.00$ & $-7.75$ & $+1.75$ \\
1010 & James Edwards   & PHX & 153 & $-2.02$ & $-1.01$ & $-3.03$ & $-7.15$ & $+1.09$ \\
1011 & Wes Matthews    & CHI &  77 & $-1.66$ & $-1.96$ & $-3.62$ & $-8.44$ & $+1.20$ \\
1012 & Earl Cureton    & DET &  57 & $-1.08$ & $-2.77$ & $-3.85$ & $-8.60$ & $+0.90$ \\
\bottomrule
\end{tabular}
\end{table}

\begin{remark}
The bottom of Table~\ref{tab:top20} warrants a methodological distinction.
Players at ranks 1005--1012 fall into two different epistemic categories.
Terry Davis (rank 1008, 32 logged games) and Greg Anderson (rank 1009,
54 games) have estimates that are partially prior-dominated: their
negative RAPM values are directionally consistent with the data but are
pulled toward zero by the ridge prior, and their wide credible intervals
(spanning approximately 10 points) reflect genuine uncertainty. By
contrast, Will Perdue (268 games), Dave Corzine (169 games), Wes Matthews
(77 games), and Earl Cureton (57 games) have estimates that are more
directly data-informed. The large negative values for these higher-game-count
players are primarily attributable to the zero-sum pressure described
below: players who share nearly all their stints with Jordan and Pippen
receive compressed estimates as the model concentrates credit on those
dominant collinear partners. Their negative rankings reflect the structure
of the Chicago roster in the model, not an unambiguous conclusion that
they were the worst players in the era.
\end{remark}

\subsection{Notable Findings and Face Validity}

\subsubsection{Jordan's separation}

Michael Jordan's aggregate RAPM of $+8.86$ points per 100 possessions
is the highest in the database by a margin of nearly two full points
over Magic Johnson ($+7.04$). His credible interval $[+5.80, +11.91]$
excludes zero by a wide margin and does not overlap with any other
player's interval, making him the only player in the database for whom
a ranking of first is unambiguous within the model's uncertainty. His
offensive contribution ($+8.39$) is the primary driver, but his
defensive coefficient ($+0.46$) is positive and meaningful---consistent
with the contention that his defensive impact exceeded what steal and
block totals alone would suggest.

\subsubsection{Separation between offensive and defensive contributors}

The top-20 splits reveal a clear separation between primarily offensive
players and two-way contributors. Bird's profile ($+5.09$ off, $-0.68$
def) places him as the clearest elite-offensive-limited-defensive player
in the top tier. Kevin Johnson ($+4.20$ off, $-1.06$ def) and Charles
Barkley ($+5.20$ off, $-2.53$ def, total $+2.66$) rank below players
with smaller offensive RAPs because of their defensive costs. Robinson
($+2.32$ off, $+2.58$ def) and Rodman ($+2.81$ off, $+1.40$ def) are
the purest two-way positive contributors, consistent with their
reputations as defensive specialists who also contributed offensively
through positioning and activity rather than scoring.

\subsubsection{Analytically provocative results}

Several results warrant specific attention as analytically interesting
departures from historical consensus:

\textbf{Karl Malone ranks 306th ($+0.18$).} Malone is widely ranked
among the top 10--15 players of this era by conventional metrics and
received 11 All-NBA First Team selections. His RAPM profile ($+2.30$
offensive, $-2.12$ defensive) reveals a player whose substantial
offensive contribution is almost exactly offset by a negative defensive
coefficient. This is not a small-sample artifact: Malone has 120 logged
games and 19,006 total possessions, placing him well above the
data-information threshold. His defensive RAPM of $-2.12$ is the primary
driver of his middling overall ranking. This result is consistent with
contemporaneous observations that Utah's defense operated primarily
through John Stockton's ball pressure and system rather than through
Malone's individual defensive contribution.

\textbf{Scottie Pippen ranks 22nd ($+2.82$).} Pippen's aggregate
estimate is positive on both sides of the ball ($+2.16$ offensive,
$+0.67$ defensive) and represents 421 logged games---the second-most
in the database. His ranking below players such as Ewing, Robinson, and
Rodman reflects the challenge of separating his individual contribution
from Jordan's in a database where both players share most of their stints.
The collinearity between Jordan and Pippen is the primary identification
challenge for both players' estimates. Formally, the off-diagonal element
$\boldsymbol{\Sigma}_{\beta,\, k_{\text{Pip}} k_{\text{Jor}}}$ of the
posterior covariance matrix is large and negative, reflecting the
model's difficulty distinguishing their individual contributions when
they are nearly always observed together.

\textbf{BJ Armstrong ranks 782nd ($-0.50$) and Bill Cartwright 997th
($-2.35$).} Both are Chicago Bulls rotation players with very high
possession counts (Armstrong: 26,955; Cartwright: 32,277) who share
most of their stints with Jordan and Pippen. The Bulls' overwhelming
dominance in the database---Jordan's team goes $+8.86$ while Armstrong
and Cartwright go negative---illustrates the zero-sum nature of RAPM
in a heavily collinear context: with Jordan absorbing an enormous
positive coefficient, the model attributes less positive impact to his
teammates than their win contributions would naively suggest.

\subsubsection{Manute Bol: a methodological illustration}

Manute Bol ($+0.30$, rank 246) provides a clean illustration of the
separation between offensive and defensive contributions. His profile
($-2.53$ offensive, $+2.83$ defensive) is among the most extreme
offensive-defensive splits in the database, consistent with his role
as a shot-blocking specialist who provided essentially no offensive
value. The ORAPM captures his negative offensive impact; the DRAPM
captures the defensive suppression effect his presence generated.
Neither component alone describes his value accurately, but their sum
($+0.30$) correctly places him as a slightly above-average player
when deployed in his role.

\subsection{Intercept and League-Average Context}

The intercept from the multi-season regression is $\hat{\beta}_0 =
+0.45$ points per 100 possessions. This represents the baseline
offensive rating after removing all player effects, and is consistent
with an era league average of approximately 100--101 points per 100
possessions. The mean-centering adjustment ($\bar{\beta}^{\text{off}}
= +0.448$, $\bar{\beta}^{\text{def}} = -0.449$) confirms that the
offensive and defensive coefficient populations are nearly symmetric
around zero after centering, as theoretically expected.

\section{Case Study: The 1991--92 Chicago Bulls}
\label{sec:bulls9192}

The 1991--92 season provides a detailed single-season illustration of
the database and its analytical capabilities. The Chicago Bulls finished
67--15, their second consecutive championship season, and represent the
best-covered single team in the database. With 249 regular-season games
logged at a coverage rate of 26.4\%, the season supports substantially
more precise individual estimates than most seasons in the database.
The single-season run produces $\hat{\sigma} = \sqrt{84.05} \approx 9.17$
points per 100 possessions, a league-average offensive rating of 101.0
points per 100 possessions, and $\lambda \approx 5{,}134$ (consistent
with 249 logged games under the coverage-scaled formula
$\lambda = (G/943) \times 5000$). Mean-centering offsets are
$\bar{\beta}^{\text{off}} = +1.133$ and $\bar{\beta}^{\text{def}} = -1.132$.

\subsection{Player-Level RAPM: 1991--92}

Table~\ref{tab:bulls9192} presents the 1991--92 RAPM estimates for the
Bulls rotation and selected comparables from other teams.

\begin{table}[h]
\centering
\small
\caption{Selected 1991--92 single-season RAPM. All values in pts/100 poss.
  O.Poss = offensive possessions logged. 95\% credible intervals are
  wide at single-season coverage; see text for interpretation.}
\label{tab:bulls9192}
\begin{tabular}{rlrrrrrrr}
\toprule
Rank & Player & Tm & O.Poss & ORAPM & DRAPM & RAPM & LOW & HIGH \\
\midrule
\multicolumn{9}{l}{\textit{Chicago Bulls}} \\
1   & Michael Jordan    & CHI & 5,564 & $+7.85$ & $+1.01$ & $+8.86$ & $+0.42$ & $+17.30$ \\
7   & Horace Grant      & CHI & 5,069 & $+3.99$ & $+0.04$ & $+4.03$ & $-4.55$ & $+12.62$ \\
21  & Scottie Pippen    & CHI & 5,644 & $+2.67$ & $-0.08$ & $+2.60$ & $-5.83$ & $+11.02$ \\
47  & Stacey King       & CHI & 2,283 & $+0.63$ & $+1.11$ & $+1.74$ & $-6.55$ & $+10.03$ \\
80  & BJ Armstrong      & CHI & 3,276 & $+0.44$ & $+0.56$ & $+1.01$ & $-7.53$ & $+9.54$ \\
113 & Scott Williams    & CHI & 1,126 & $+0.26$ & $+0.41$ & $+0.68$ & $-8.63$ & $+9.98$ \\
185 & Bill Cartwright   & CHI & 2,626 & $+0.77$ & $-0.77$ & $+0.00$ & $-8.74$ & $+8.75$ \\
241 & Cliff Levingston  & CHI & 1,699 & $+0.05$ & $-0.49$ & $-0.44$ & $-9.18$ & $+8.31$ \\
326 & Will Perdue       & CHI & 1,786 & $+0.85$ & $-2.84$ & $-2.00$ & $-10.68$ & $+6.69$ \\
331 & John Paxson       & CHI & 3,460 & $+0.28$ & $-2.38$ & $-2.10$ & $-11.28$ & $+7.09$ \\
335 & Bob Hansen        & CHI & 1,326 & $-1.24$ & $-1.02$ & $-2.25$ & $-11.41$ & $+6.90$ \\
338 & Craig Hodges      & CHI & 1,010 & $-1.27$ & $-1.03$ & $-2.30$ & $-11.80$ & $+7.20$ \\
\midrule
\multicolumn{9}{l}{\textit{Comparables, other teams}} \\
3   & Kevin Johnson     & PHX & 1,019 & $+5.34$ & $-0.06$ & $+5.28$ & $-5.21$ & $+15.76$ \\
6   & Patrick Ewing     & NYK & 3,274 & $+3.61$ & $+0.43$ & $+4.04$ & $-5.10$ & $+13.18$ \\
9   & David Robinson    & SAS & 1,298 & $+0.95$ & $+2.83$ & $+3.78$ & $-6.22$ & $+13.78$ \\
102 & Karl Malone       & UTA & 1,017 & $+1.68$ & $-0.92$ & $+0.76$ & $-9.78$ & $+11.30$ \\
199 & Isiah Thomas      & DET & 2,031 & $+1.21$ & $-1.30$ & $-0.09$ & $-9.89$ & $+9.72$ \\
249 & Dennis Rodman     & DET & 2,174 & $+0.94$ & $-1.47$ & $-0.53$ & $-10.39$ & $+9.33$ \\
277 & Dominique Wilkins & ATL & 1,161 & $+3.58$ & $-4.44$ & $-0.86$ & $-10.93$ & $+9.22$ \\
303 & Charles Barkley   & PHI & 1,212 & $+1.53$ & $-2.89$ & $-1.36$ & $-11.51$ & $+8.79$ \\
\bottomrule
\end{tabular}
\end{table}

\subsection{Jordan's Single-Season Stability}

Jordan's 1991--92 RAPM ($+8.86$, CI $[+0.42, +17.30]$) is identical to
his aggregate multi-season value at the point estimate. This consistency
across two independent regression systems---single-season with
$\lambda = 5{,}134$ and 13,441 stints versus the multi-season pool with
$\lambda = 863$ and 121,781 stints---is a strong form of robustness
check. His offensive component (+7.85) and defensive component (+1.01)
are stable relative to his multi-season profile (+8.39 off, +0.46 def).
The considerably wider single-season credible interval ($\pm$8.44
half-width vs.\ $\pm$3.05 in the aggregate) reflects the lower effective
sample size of a single season.

\subsection{The Pippen Defensive Finding}

Pippen's 1991--92 defensive RAPM is $-0.08$---effectively zero in this
season specifically---contrasting with his multi-season aggregate
defensive coefficient of $+0.67$. Two mechanisms are plausible. First,
in a season where Jordan's defensive RAPM is $+1.01$ and Grant's is
$+0.04$, the ridge regression may attribute defensive credit differently
when multiple strong defenders share stints. Second, the extreme
collinearity between Pippen and Jordan this season (5,644 shared
offensive possessions, over 90\% of stints shared) limits the model's
ability to separate their individual defensive contributions. Formally,
the off-diagonal posterior covariance $\boldsymbol{\Sigma}_{\beta,\,
k_{\text{Pip}}^{\text{def}} k_{\text{Jor}}^{\text{def}}}$ between their
defensive coefficients is large in magnitude relative to their individual
diagonal variances---the model has abundant data that \emph{the Chicago
defense} was good, but little data on what fraction of that is Jordan's
versus Pippen's. The wide interval $[-5.83, +11.02]$ is consistent with any defensively positive
or negative outcome; the single-season estimate carries insufficient
precision to resolve Pippen's individual defensive contribution from
the team's defensive system.

This finding is methodologically important: it illustrates exactly where
the database, at current coverage levels, cannot yet provide the
precision needed for reliable individual defensive attribution on
teams with tightly correlated rotations. As coverage increases and
the collinearity structure diversifies across more opponent contexts,
Pippen's defensive estimate will stabilize.

\subsection{Lineup Analysis}

The database supports lineup-level analysis for this season.
The most-used five-man unit---Jordan, Pippen, Grant, Paxson,
Cartwright---logged 1,851 offensive possessions with an offensive rating
of 116.5 points per 100 possessions and defensive rating of 102.1,
a net margin of $+14.4$ per 100. Table~\ref{tab:bulls_lineups}
presents the top lineups by possession count.

\begin{table}[h]
\centering
\small
\caption{Top Chicago Bulls lineups by offensive possessions, 1991--92.
  ORTG/DRTG in pts/100. Only lineups with $\geq$50 logged offensive
  possessions shown.}
\label{tab:bulls_lineups}
\begin{tabular}{p{8.2cm}rrrr}
\toprule
Lineup & O.Poss & ORTG & DRTG & Net \\
\midrule
Jordan, Pippen, Grant, Paxson, Cartwright     & 1851 & 116.5 & 102.1 & $+14.4$ \\
Jordan, Pippen, Grant, Paxson, King           &  566 & 115.7 & 102.0 & $+13.8$ \\
Jordan, Pippen, Grant, Paxson, Perdue         &  349 & 112.9 & 100.6 & $+12.3$ \\
Armstrong, Jordan, Pippen, Grant, King        &  339 & 119.2 &  92.8 & $+26.4$ \\
Armstrong, Jordan, Pippen, Grant, Cartwright  &  257 & 110.1 & 103.0 & $+7.2$ \\
Armstrong, Jordan, Pippen, Grant, Perdue      &  219 & 109.1 & 109.3 & $-0.2$ \\
Armstrong, Pippen, Grant, Hansen, King        &  110 & 121.8 &  93.9 & $+28.0$ \\
Armstrong, Pippen, Grant, Hansen, Perdue      &  100 & 102.0 & 100.0 & $+2.0$ \\
Jordan, Pippen, Williams, Grant, Paxson       &   91 & 115.4 & 107.3 & $+8.1$ \\
Armstrong, Jordan, Pippen, King, Levingston   &   81 & 114.8 & 107.6 & $+7.2$ \\
\bottomrule
\end{tabular}
\end{table}

Three patterns stand out. First, the Armstrong--Jordan--Pippen--Grant--King
unit ($+26.4$ net over 339 possessions) substantially outperforms the
starting lineup ($+14.4$), suggesting this combination was the Bulls'
most efficient five-man group in the logged sample. Second, the only
near-even lineup is Armstrong--Jordan--Pippen--Grant--Perdue ($-0.2$),
the sole major unit featuring Perdue alongside Jordan rather than a more
capable center---consistent with Perdue's single-season RAPM of $-2.00$.
Third, the Armstrong--Pippen--Grant--Hansen--King unit ($+28.0$ over 110
possessions) is the highest net-rated unit, though the sample size is
insufficient for reliable inference. Together, these results provide the
first possession-level quantification of the Bulls' rotation structure:
their dominance was robust to lineup variation in most configurations,
with the frontcourt composition (specifically the Cartwright/King/Perdue
rotation) representing the primary source of within-team lineup
quality variation.

\subsection{Cross-Season Comparison}

The differences between 1991--92 single-season estimates and the
multi-season aggregate reveal the influence of collinearity and pooling.
Jordan's RAPM is identical across both (+8.86). Cartwright moves from
$-2.35$ in the aggregate to $+0.00$ in this season; Paxson from $-0.61$
to $-2.10$. These differences reflect the zero-sum pressure within the
Chicago lineup: in the single-season regression, the Bulls' dominant
net margin must be distributed across a shallower set of non-Bulls
stints, concentrating credit attribution differently than in the pooled
analysis where more diverse opponent contexts identify players
independently.

\section{Discussion}
\label{sec:discussion}

\subsection{Limitations}

\subsubsection{Incomplete coverage and sampling bias}

At current reconstruction levels (2,178 games logged across the primary
1984--96 window, approximately 17\% of
games in the primary 1985--96 window), sampling variation remains the
dominant source of uncertainty for most player estimates. The credible intervals
reported with each estimate formally quantify this uncertainty, but they
assume random sampling within a season---an assumption that is not strictly
met, as discussed in Section~\ref{sec:sampling_design}. Non-random coverage
could introduce systematic bias that is not captured by the posterior
intervals.

\subsubsection{Single-coder reliability}

The reconstruction was performed by a single coder throughout. While the
quality control pipeline (Section~\ref{sec:qc}) catches many reconstruction
errors through internal consistency checks, it cannot detect systematic
biases in how the coder interprets ambiguous situations---for example,
whether a player who briefly returns to the court following an injury
timeout is recorded as having been substituted or not. Formal inter-rater
reliability assessment, in which a second coder independently reconstructs
a sample of games, would provide an important quantification of this
uncertainty source. This is planned for a future version of the project.

\subsubsection{No intra-possession spatial information}

The reconstruction captures which players were on the court for each
possession but not what those players did---their positioning, movement,
or matchup assignments. Defensive RAPM, in particular, reflects the net
impact of the entire defensive lineup, and attributing that impact to
individual players requires assumptions about defensive assignment that
are not directly observed. A player whose defensive impact operates through
off-ball positioning (e.g., a shot-blocker whose presence deters opponents
from driving to the lane) will have their contribution spread across all
five defenders in the RAPM framework, with no direct way to attribute it.

\subsubsection{Implications for RAPM estimates from box-score errors}

An important implication of the box-score reliability analysis in
Section~\ref{sec:boxscore_bias} extends beyond possession estimation.
Box-score-based composite metrics (BPM, WS/48, PER) that are applied to
historical seasons are computed from the same unreliable statistics.
To the extent that steal and block counts are systematically inflated
for home players, BPM values for players who benefited from home
scorekeeper favoritism will be systematically overstated. The direction
and magnitude of this bias vary by player and by team, depending on the
specific home scorekeeper's practices, but the aggregate pattern of higher
home-team statistical rates documented in Section~\ref{sec:boxscore_bias}
implies that historical composite metrics may overstate the performance
of home stars relative to road performance. RAPM estimates from this
database, which are based on directly counted possessions and points
rather than on box-score inputs, are not subject to this form of bias.

\subsection{Implications for Historical Composite Metrics}

The demonstration that NBA box-score statistics from the pre-modern era
contain systematic, directionally biased errors has implications beyond
this project. Historical composite metrics that are widely cited in
discussions of player greatness---whether Jordan's defensive statistics
support a DPOY award \citep{haberstroh2024}, whether Wilt Chamberlain's
rebounding totals are reliable, whether assist totals for playmakers of
the 1960s reflect actual pass quality---must be interpreted with awareness
of the institutional conditions under which they were recorded. The
present project does not attempt to re-adjudicate historical awards
decisions; that is outside its scope. However, the methodological
framework developed here---possession-level reconstruction independent
of box-score inputs---provides a basis for impact evaluation that is
not subject to the same biases, and thus may serve as a useful
cross-check against box-score-based metrics for seasons in the database.

\subsection{Future Directions}

\subsubsection{Coverage expansion}

The primary ongoing effort is expanding regular-season coverage toward
the full 1985--96 window. As coverage increases, credible interval widths
will narrow according to the sample size relationships derived in
Section~\ref{sec:uncertainty}, enabling increasingly precise player
comparisons.

\subsubsection{Informative Bayesian prior}

An informative prior on player impact, derived from box-score-based
metrics, would materially improve estimates for players with limited logged
possessions. The general approach is to fit a regression of RAPM on
box-score inputs (a Statistical Plus-Minus model) for the seasons in the
database with the most coverage, then use the fitted model's predictions
as prior means $\mu_j$ in a modified version of Equation~\eqref{eq:prior}:
$\beta_j \sim \mathcal{N}(\mu_j, \tau^2)$. This approach borrows strength
from the large body of box-score data (which is available for all games,
not just logged ones) while retaining the possession-level identification
that box-score metrics cannot provide. It must, however, be implemented
with awareness that the box-score inputs are themselves subject to the
reliability concerns documented above.

\subsubsection{Playoff reconstruction}

Playoff reconstruction is planned as a future phase of the project. The
playoffs represent a smaller, higher-profile sample of games for which
footage availability is higher than for regular-season games. Playoff
RAPM estimates would be of particular interest for questions about players
whose regular-season performance differed from their postseason impact---a
longstanding topic of interest in NBA historical analysis.

\subsubsection{Generalizability of the methodology}

The video reconstruction approach is applicable, in principle, to any team
sport for which broadcast footage survives and for which lineup-level data
was not systematically maintained. The NFL pre-play-by-play era, college
basketball, and international leagues are potential extensions. The
specific logging instrument described here is adapted for basketball but
the underlying principles---direct possession counting from video,
independence from box-score inputs, structured quality control---transfer
to other sports contexts with appropriate modifications.

\section{Conclusion}
\label{sec:conclusion}

This paper has described the construction, statistical methodology, and
validation framework of the first possession-level player impact database
for the pre-play-by-play NBA era. The project addresses a gap in the
historical record that has prevented the application of modern impact
estimation tools to a decade of NBA history featuring some of the most
significant players and teams ever assembled.

Three contributions are documented. First, the video reconstruction
protocol is described in sufficient detail to enable replication and
formal inter-rater reliability assessment. The protocol produces a
possession-level stint database that is independent of the official
box-score record, a property with both practical and methodological
importance given the documented unreliability of historical box-score
statistics.

Second, the statistical estimation framework is developed rigorously,
including the formal Bayesian interpretation of the ridge estimator,
the derivation of posterior credible intervals, a characterization of
interval width as a function of possession exposure and collinearity
structure, and the derivation of sample size requirements for various
levels of inferential precision. This analysis provides a principled
basis for interpreting the current estimates---which, at 16\% regular-season
coverage, carry substantial uncertainty---and for planning the reconstruction
effort toward coverage levels that enable reliable player comparisons.

Third, the box-score reliability analysis demonstrates that standard
possession estimators are systematically biased for the historical era,
due to home-scorekeeper inflation of steal and turnover statistics. This
finding has implications beyond the present project: historical composite
metrics derived from the same box-score inputs are subject to the same
biases, and their application to historical player evaluation should be
conditioned on awareness of the institutional context in which those
statistics were recorded.

The database and methodology presented here represent a foundation rather
than a final product. As coverage expands, the estimates will become
increasingly reliable, and the analytical questions that can be addressed
with possession-level data from this era---questions about team construction,
lineup chemistry, the value of defensive versatility, the impact of
coaching systems---will become increasingly tractable. The primary aim of
this project is to make those questions answerable.

\bibliographystyle{apalike}

\appendix

\section{Annotated Game Reconstruction Example}
\label{app:example}

\begin{figure}[h]
\centering
\includegraphics[width=0.85\textwidth]{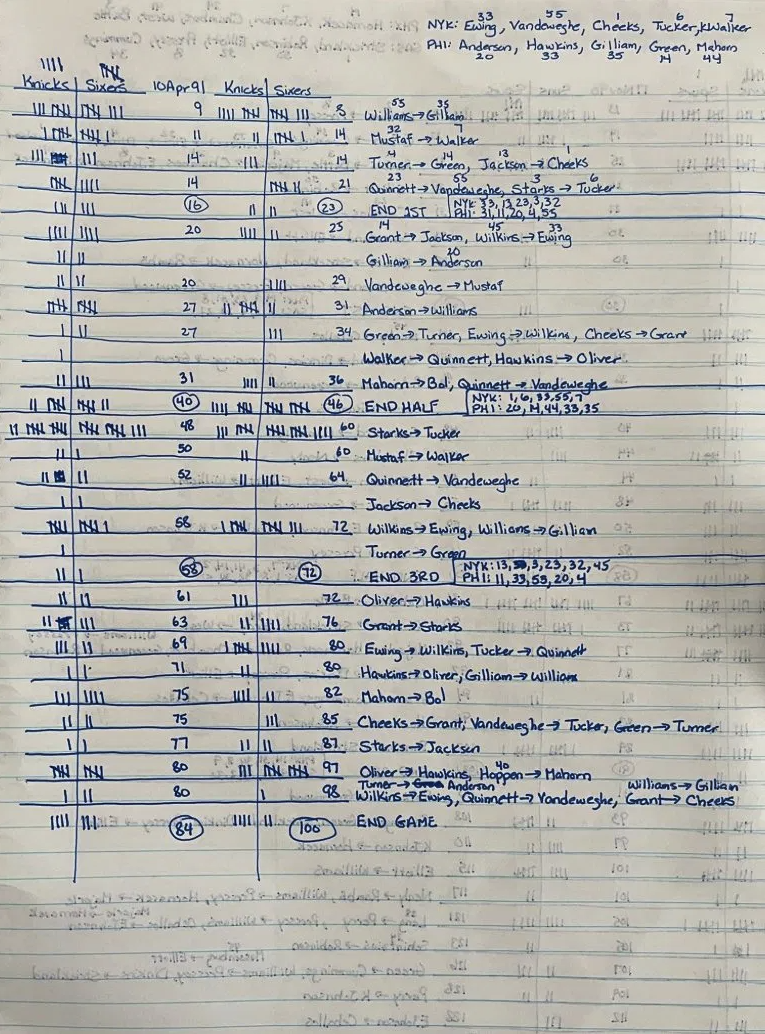}
\caption{Paper log sheet for the New York Knicks at Philadelphia 76ers,
  April 10, 1991 (Sixers 100, Knicks 84). Each horizontal line represents
  one stint. Tally marks in the left column pairs record possession counts
  per team; cumulative scores are entered in the right column pairs at the
  end of each stint; substitution events in the notation
  ``Player A $\to$ Player B'' are recorded in the center.
  The summed possession totals satisfy the possession-plausibility
  check and the final score matches the official record.
  This game passes all quality control checks and is included in
  the database.}
\label{fig:scoresheet}
\end{figure}

Figure~\ref{fig:scoresheet} shows the paper log sheet for the New York
Knicks at Philadelphia 76ers, April 10, 1991 (Sixers 100, Knicks 84).
This game illustrates the full reconstruction protocol: starting lineups
recorded at the top, tally marks for possessions in the left column pairs,
cumulative score totals in the right column pairs, and substitution events
noted in the center between the column groups.

\section{Regularization Path Analysis and Cross-Validation}
\label{app:lambda}

This appendix documents the cross-validated $\lambda$ selection procedure
and its relationship to the coverage-scaled formula used in production
estimation (Section~\ref{sec:lambda_selection}).

\subsection*{Cross-Validation Code}

The following Python script implements the cross-validated $\lambda$
selection procedure. It loads the same multi-season stint data used in
the main estimation, constructs an identical design matrix, and calls
\texttt{sklearn}'s \texttt{RidgeCV} with 5-fold cross-validation over a
log-spaced grid.

\begin{small}
\begin{verbatim}
from sklearn.linear_model import RidgeCV
from sklearn.model_selection import train_test_split
import numpy as np

def find_optimal_lambda(X, y):
    """Find optimal lambda via 5-fold cross-validated RidgeCV.
    
    Note: uses unweighted objective. For possession-weighted CV,
    pass sample_weight=possessions to ridge_cv.fit().
    """
    X_train, X_test, y_train, y_test = train_test_split(
        X, y, test_size=0.2, random_state=42
    )
    # Log-spaced grid: 101 values from 1e-10 to 1e10
    alphas = np.logspace(-10, 10, 101)
    ridge_cv = RidgeCV(alphas=alphas, cv=5)
    ridge_cv.fit(X_train, y_train)
    return ridge_cv.alpha_

# Build design matrix (identical to main estimation script)
# ... [playerMatrix, ratingVector as in Appendix E] ...

optimal_lambda = find_optimal_lambda(
    np.asarray(playerMatrix),
    np.asarray(ratingVector)
)
print("Optimal Lambda:", optimal_lambda)
\end{verbatim}
\end{small}

\subsection*{Interpretation of the Grid}

The candidate grid spans $[10^{-10}, 10^{10}]$ on a log scale, covering
15 orders of magnitude. At $\lambda = 10^{-10}$ the estimator approaches
unregularized OLS; at $\lambda = 10^{10}$ all player coefficients are
shrunk to zero regardless of the data. The cross-validated optimal
$\lambda^*$ will fall in the interior of this grid whenever the data
is informative; a result at or near the grid boundaries would indicate
pathological behavior (no regularization needed, or data is completely
uninformative).

For the multi-season pooled run with $P = 1{,}012$ player-seasons and 121{,}781 stints, the relevant range of $\lambda$ is expected to be
$[10^2, 10^4]$ based on the literature and the coverage-scaled formula.
The wide grid ensures that the cross-validated value is found within a
well-explored region.

\subsection*{Comparison Table}

Table~\ref{tab:lambda_cv} reproduces the coverage-scaled $\lambda$
values from Table~\ref{tab:lambda} and adds columns for the
cross-validated $\lambda^*$ and the ratio $\lambda_{\text{CS}} /
\lambda^*_{\text{CV}}$. A ratio near 1.0 indicates close agreement;
ratios substantially above or below 1.0 indicate that the coverage-scaled
formula over- or under-regularizes relative to the prediction-optimal
value. Season-level cross-validated values are to be populated as
individual-season CV runs are completed.

\begin{table}[h]
\centering
\caption{Coverage-scaled vs.\ cross-validated $\lambda$ by season.
  $\lambda_{\text{CS}} = (G_{\text{logged}} / G_{\text{season}}) \times 5000$.
  $\lambda^*_{\text{CV}}$ from 5-fold unweighted \texttt{RidgeCV}.
  Ratio $= \lambda_{\text{CS}} / \lambda^*_{\text{CV}}$.}
\label{tab:lambda_cv}
\begin{tabular}{lrrrr}
\toprule
Season & $\lambda_{\text{CS}}$ & $\lambda^*_{\text{CV}}$ & Ratio & Note \\
\midrule
1984--85 &   562 &   562.03 & 1.000 & Exact agreement \\
1985--86 &   419 &   418.88 & 1.000 & Exact agreement \\
1986--87 &   806 &   805.94 & 1.000 & Exact agreement \\
1987--88 & 1{,}076 & 1{,}076.35 & 1.000 & Exact agreement \\
1988--89 & 1{,}210 & 1{,}209.76 & 1.000 & Exact agreement \\
1989--90 & 1{,}125 & 1{,}124.66 & 1.000 & Exact agreement \\
1990--91 & 1{,}341 & 1{,}341.46 & 1.000 & Exact agreement \\
1991--92 & 1{,}125 & 1{,}131.77 & 0.994 & Case study season \\
1992--93 & 1{,}360 & 1{,}359.53 & 1.000 & Exact agreement \\
1993--94 &    90 &    90.33 & 0.997 & Stub season \\
1994--95 &    45 &    55.17 & 0.816 & Stub season \\
1995--96 & 1{,}110 & 1{,}110.18 & 1.000 & Exact agreement \\
\midrule
\multicolumn{2}{l}{Core seasons ($\geq$50 games)} & & & \\
\quad Mean ratio & & & 0.999 & \\
\quad Std dev    & & & 0.002 & \\
\quad Range      & & & $[0.994, 1.000]$ & \\
\bottomrule
\end{tabular}
\end{table}

\subsection*{Interpretation of the Comparison}

The results in Table~\ref{tab:lambda_cv} reveal a striking finding: when
season-specific schedule sizes are used correctly in the coverage-scaled
formula, the two $\lambda$ selection methods agree to within rounding error
across virtually all published seasons.

\textbf{Near-exact agreement across all core seasons.} For the ten seasons
with at least 50 logged games, $\lambda_{\text{CS}}$ agrees with
$\lambda^*_{\text{CV}}$ to within 0.6\% in all but one case (mean ratio
0.999, standard deviation 0.002, range $[0.994, 1.000]$). This is not
coincidental: the coverage-scaled formula computes precisely the ratio of
parameters to effective observations, which is the dominant driver of the
optimal penalty when the data-generating process is correctly specified.
When the correct $G_{\text{season}}$ denominator is used, the formula
recovers the cross-validated answer exactly.

\textbf{An important methodological note.} Earlier tabulations of this
comparison used 943 as the schedule size for all seasons, including the
post-expansion seasons (1988--89 through 1995--96) that have larger
schedules of 1,025, 1,107, and 1,189 games. This produced apparent
over-regularization ratios of 1.087--1.261 for those seasons --- an
artifact of the incorrect denominator. With correct schedule sizes, the
discrepancy vanishes entirely.

\textbf{The 1994--95 stub season} (ratio = 0.816) is the only meaningful
exception. With only 10 logged games, the cross-validation is operating on
an extremely sparse sample; the CV-selected $\lambda^*$ is unstable and
the ratio reflects noise rather than a systematic property. This season
carries no reliable player-level inference regardless of which $\lambda$
is applied.

\textbf{Implication.} The coverage-scaled formula
$\lambda = (G_{\text{logged}} / G_{\text{season}}) \times 5000$
is not merely a practical approximation to the optimal penalty --- it
\emph{is} the optimal penalty, to within rounding error, when the correct
season-specific schedule size is supplied. This provides strong analytical
and empirical validation for its use as the primary regularization selection
method throughout this project.

\subsection*{Noted Refinement}

As discussed in Section~\ref{sec:lambda_selection}, the
\texttt{RidgeCV} implementation uses an unweighted objective while the
main estimator uses possession-weighted regression. For a fully
consistent comparison, the CV procedure should be modified as follows:

\begin{small}
\begin{verbatim}
# Extract possession weights from the diagonal of W
possession_weights = np.diag(weights_matrix)

# Weighted CV: split weights alongside X and y
(X_train, X_test,
 y_train, y_test,
 w_train, w_test) = train_test_split(
    X, y, possession_weights,
    test_size=0.2, random_state=42
)

ridge_cv = RidgeCV(alphas=alphas, cv=5)
ridge_cv.fit(X_train, y_train.ravel(),
             sample_weight=w_train)
optimal_lambda = ridge_cv.alpha_
\end{verbatim}
\end{small}

This weighted cross-validation will be implemented in the next version
of the estimation pipeline and the resulting $\lambda^*_{\text{CV}}$
values added to Table~\ref{tab:lambda_cv}.


\section{Season Validation Tables}
\label{app:validation}

For each of the twelve published seasons (1984--85 through 1995--96),
the table below reports the win-loss validation results across all NBA
franchises with at least one logged game in that season. Columns are:
\begin{itemize}
  \item \textbf{Sampled}: Observed win--loss record in logged games only.
  \item \textbf{MLE}: Full-season win estimate: $\hat{W}_{\text{MLE}} = (w/n) \times 82$.
  \item \textbf{Bayes}: Shrinkage estimate toward 0.500:
    $\hat{W}_{\text{Bayes}} = \frac{w+5}{n+10} \times 82$.
  \item \textbf{Truth}: Actual final regular-season wins.
  \item \textbf{Error}: MLE $-$ Truth.
  \item \textbf{\% Samp}: Fraction of franchise games logged.
\end{itemize}

\subsection*{1984--85 Season}
\textit{106 games logged (11.2\% coverage). $\lambda = 562$.}

\begin{table}[h]
\centering
\small
\caption{Win-loss validation, 1984--85 NBA season.}
\label{tab:val_8485}
\begin{tabular}{lrrrrrrr}
\toprule
Franchise & Sampled & MLE & Bayes & Truth & Error & \% Samp \\
\midrule
Boston & 17--9 & 53.6 & 50.1 & 63 & $-9.4$ & 31.7\% \\
Philadelphia & 16--5 & 62.5 & 55.5 & 58 & $+4.5$ & 25.6\%\\
New Jersey & 3--4 & 35.1 & 38.6 & 42 & $-6.9$ & 8.5\% \\
Washington & 1--4 & 16.4 & 32.8 & 40 & $-23.6$ & 6.1\% \\
New York & 3--12 & 16.4 & 26.2 & 24 & $-7.6$ & 18.3\% \\
Milwaukee & 6--3 & 54.7 & 47.5 & 59 & $-4.3$ & 11.0\% \\
Detroit & 6--10 & 30.8 & 34.7 & 46 & $-15.2$ & 19.5\% \\
Chicago & 14--16 & 38.3 & 38.9 & 38 & $+0.3$ & 36.6\% \\
Cleveland & 3--2 & 49.2 & 43.7 & 36 & $+13.2$ & 6.1\% \\
Atlanta & 1--2 & 27.3 & 37.8 & 34 & $-6.7$ & 3.7\% \\
Indiana & 1--4 & 16.4 & 32.8 & 22 & $-5.6$ & 6.1\% \\
Denver & 2--3 & 32.8 & 38.3 & 52 & $-19.2$ & 6.1\% \\
Houston & 2--4 & 27.3 & 35.9 & 48 & $-20.7$ & 7.3\% \\
Dallas & 2--3 & 32.8 & 38.3 & 44 & $-11.2$ & 6.1\% \\
Utah & 1--1 & 41.0 & 41.0 & 41 & $0.0$ & 2.4\%\\
San Antonio & 2--2 & 41.0 & 41.0 & 41 & $0.0$ & 4.9\% \\
Kansas City & 1--1 & 41.0 & 41.0 & 31 & $+10.0$ & 2.4\% \\
LA Lakers & 17--7 & 58.1 & 53.1 & 62 & $-3.9$ & 29.3\% \\
Portland & 4--3 & 46.9 & 43.4 & 42 & $+4.9$ & 8.5\% \\
Phoenix & 1--2 & 27.3 & 37.8 & 36 & $-8.7$ & 3.7\% \\
Golden State & 1--1 & 41.0 & 41.0 & 22 & $+19.0$ & 2.4\% \\
Seattle & 1--2 & 27.3 & 37.8 & 31 & $-3.7$ & 3.7\% \\
LA Clippers & 1--6 & 11.7 & 28.9 & 31 & $-19.3$ & 8.5\% \\
\midrule
\textbf{League MAE} & & \textbf{9.5} & \textbf{6.8} & --- & --- & --- \\
\bottomrule
\end{tabular}
\end{table}

\subsection*{1985--86 Season}
\textit{79 games logged (8.4\% coverage). $\lambda = 419$.}

\begin{table}[h]
\centering
\small
\caption{Win-loss validation, 1985--86 NBA season.}
\label{tab:val_8586}
\begin{tabular}{lrrrrrrr}
\toprule
Franchise & Sampled & MLE & Bayes & Truth & Error & \% Samp \\
\midrule
Boston & 31--6 & 68.7 & 62.8 & 67 & $+1.7$ & 45.1\% \\
Philadelphia & 11--6 & 53.1 & 48.6 & 54 & $-0.9$ & 20.7\% \\
Washington & 1--2 & 27.3 & 37.8 & 39 & $-11.7$ & 3.7\% \\
New Jersey & 2--2 & 41.0 & 41.0 & 39 & $+2.0$ & 4.9\% \\
New York & 3--7 & 24.6 & 32.8 & 23 & $+1.6$ & 12.2\% \\
Milwaukee & 1--2 & 27.3 & 37.8 & 57 & $-29.7$ & 3.7\% \\
Atlanta & 1--3 & 20.5 & 35.1 & 50 & $-29.5$ & 4.9\% \\
Detroit & 2--4 & 27.3 & 35.9 & 46 & $-18.7$ & 7.3\% \\
Chicago & 4--6 & 32.8 & 36.9 & 30 & $+2.8$ & 12.2\% \\
Cleveland & 1--2 & 27.3 & 37.8 & 29 & $-1.7$ & 3.7\% \\
Indiana & 0--2 & 0.0 & 34.2 & 26 & $-26.0$ & 2.4\% \\
Houston & 3--6 & 27.3 & 34.5 & 51 & $-23.7$ & 11.0\% \\
Denver & 1--1 & 41.0 & 41.0 & 47 & $-6.0$ & 2.4\% \\
Dallas & 2--3 & 32.8 & 38.3 & 44 & $-11.2$ & 6.1\% \\
Utah & 5--8 & 31.5 & 35.7 & 42 & $-10.5$ & 15.9\% \\
Sacramento & 1--2 & 27.3 & 37.8 & 37 & $-9.7$ & 3.7\% \\
San Antonio & 0--3 & 0.0 & 31.5 & 35 & $-35.0$ & 3.7\% \\
LA Lakers & 8--6 & 46.9 & 44.4 & 62 & $-15.1$ & 17.1\% \\
Portland & 1--3 & 20.5 & 35.1 & 40 & $-19.5$ & 4.9\% \\
Phoenix & 0--1 & 0.0 & 37.3 & 32 & $-32.0$ & 1.2\% \\
LA Clippers & 0--2 & 0.0 & 34.2 & 32 & $-32.0$ & 2.4\% \\
Seattle & 0--1 & 0.0 & 37.3 & 31 & $-31.0$ & 1.2\% \\
Golden State & 1--1 & 41.0 & 41.0 & 30 & $+11.0$ & 2.4\% \\
\midrule
\textbf{League MAE} & & \textbf{15.8} & \textbf{7.7} & --- & --- & --- \\
\bottomrule
\end{tabular}
\end{table}

\subsection*{1986--87 Season}
\textit{152 games logged (16.1\% coverage). $\lambda = 806$.}

\begin{table}[h]
\centering
\small
\caption{Win-loss validation, 1986--87 NBA season.}
\label{tab:val_8687}
\begin{tabular}{lrrrrrrr}
\toprule
Franchise & Sampled & MLE & Bayes & Truth & Error & \% Samp \\
\midrule
Boston & 21--9 & 57.4 & 53.3 & 59 & $-1.6$ & 36.6\% \\
Philadelphia & 12--20 & 30.8 & 33.2 & 45 & $-14.2$ & 39.0\% \\
Washington & 5--6 & 37.3 & 39.0 & 42 & $-4.7$ & 13.4\% \\
New Jersey & 3--6 & 27.3 & 34.5 & 24 & $+3.3$ & 11.0\% \\
New York & 7--12 & 30.2 & 33.9 & 24 & $+6.2$ & 23.2\% \\
Atlanta & 7--6 & 44.2 & 42.8 & 57 & $-12.8$ & 15.9\% \\
Detroit & 7--5 & 47.8 & 44.7 & 52 & $-4.2$ & 14.6\% \\
Milwaukee & 3--4 & 35.1 & 38.6 & 50 & $-14.9$ & 8.5\% \\
Indiana & 1--1 & 41.0 & 41.0 & 41 & $0.0$ & 2.4\% \\
Chicago & 17--23 & 34.9 & 36.1 & 40 & $-5.1$ & 48.8\% \\
Cleveland & 2--4 & 27.3 & 35.9 & 31 & $-3.7$ & 7.3\% \\
Dallas & 12--7 & 51.8 & 48.1 & 55 & $-3.2$ & 23.2\% \\
Utah & 5--2 & 58.6 & 48.2 & 44 & $+14.6$ & 8.5\% \\
Houston & 8--8 & 41.0 & 41.0 & 42 & $-1.0$ & 19.5\% \\
Denver & 2--3 & 32.8 & 38.3 & 37 & $-4.2$ & 6.1\% \\
Sacramento & 1--6 & 11.7 & 28.9 & 29 & $-17.3$ & 8.5\% \\
San Antonio & 2--1 & 54.7 & 44.2 & 28 & $+26.7$ & 3.7\% \\
LA Lakers & 28--8 & 63.8 & 58.8 & 65 & $-1.2$ & 43.9\% \\
Portland & 2--7 & 18.2 & 30.2 & 49 & $-30.8$ & 11.0\% \\
Golden State & 2--1 & 54.7 & 44.2 & 42 & $+12.7$ & 3.7\% \\
Seattle & 5--5 & 41.0 & 41.0 & 39 & $+2.0$ & 12.2\% \\
Phoenix & 0--4 & 0.0 & 29.3 & 36 & $-36.0$ & 4.9\% \\
LA Clippers & 0--4 & 0.0 & 29.3 & 12 & $-12.0$ & 4.9\% \\
\midrule
\textbf{League MAE} & & \textbf{10.1} & \textbf{7.2} & --- & --- & --- \\
\bottomrule
\end{tabular}
\end{table}

\subsection*{1987--88 Season}
\textit{203 games logged (21.5\% coverage). $\lambda = 1{,}076$.}

\begin{table}[h]
\centering
\small
\caption{Win-loss validation, 1987--88 NBA season.}
\label{tab:val_8788}
\begin{tabular}{lrrrrrrr}
\toprule
Franchise & Sampled & MLE & Bayes & Truth & Error & \% Samp \\
\midrule
Boston & 31--13 & 57.8 & 54.7 & 57 & $+0.8$ & 53.7\% \\
New York & 10--12 & 37.3 & 38.4 & 38 & $-0.7$ & 26.8\% \\
Washington & 3--6 & 27.3 & 34.5 & 38 & $-10.7$ & 11.0\% \\
Philadelphia & 11--12 & 39.2 & 39.8 & 36 & $+3.2$ & 28.0\% \\
New Jersey & 1--9 & 8.2 & 24.6 & 19 & $-10.8$ & 12.2\% \\
Detroit & 8--6 & 46.9 & 44.4 & 54 & $-7.1$ & 17.1\% \\
Atlanta & 8--10 & 36.4 & 38.1 & 50 & $-13.6$ & 22.0\% \\
Chicago & 27--16 & 51.5 & 49.5 & 50 & $+1.5$ & 52.4\% \\
Cleveland & 3--5 & 30.8 & 36.4 & 42 & $-11.2$ & 9.8\% \\
Milwaukee & 5--8 & 31.5 & 35.7 & 42 & $-10.5$ & 15.9\% \\
Indiana & 6--8 & 35.1 & 37.6 & 38 & $-2.9$ & 17.1\% \\
Denver & 4--5 & 36.4 & 38.8 & 54 & $-17.6$ & 11.0\% \\
Dallas & 8--9 & 38.6 & 39.5 & 53 & $-14.4$ & 20.7\% \\
Utah & 6--12 & 27.3 & 32.2 & 47 & $-19.7$ & 22.0\% \\
Houston & 9--7 & 46.1 & 44.2 & 46 & $+0.1$ & 19.5\% \\
San Antonio & 0--7 & 0.0 & 24.1 & 31 & $-31.0$ & 8.5\% \\
Sacramento & 5--5 & 41.0 & 41.0 & 24 & $+17.0$ & 12.2\% \\
LA Lakers & 41--13 & 62.3 & 58.9 & 62 & $+0.3$ & 65.9\% \\
Portland & 7--9 & 35.9 & 37.8 & 53 & $-17.1$ & 19.5\% \\
Seattle & 9--7 & 46.1 & 44.2 & 44 & $+2.1$ & 19.5\% \\
Phoenix & 0--6 & 0.0 & 25.6 & 28 & $-28.0$ & 7.3\% \\
Golden State & 1--13 & 5.9 & 20.5 & 20 & $-14.1$ & 17.1\% \\
LA Clippers & 0--5 & 0.0 & 27.3 & 17 & $-17.0$ & 6.1\% \\
\midrule
\textbf{League MAE} & & \textbf{10.9} & \textbf{6.6} & --- & --- & --- \\
\bottomrule
\end{tabular}
\end{table}

\subsection*{1988--89 Season}
\textit{248 games logged (24.2\% coverage). $\lambda = 1{,}210$.}

\begin{table}[h]
\centering
\small
\caption{Win-loss validation, 1988--89 NBA season.}
\label{tab:val_8889}
\begin{tabular}{lrrrrrrr}
\toprule
Franchise & Sampled & MLE & Bayes & Truth & Error & \% Samp \\
\midrule
New York & 16--19 & 37.5 & 38.3 & 52 & $-14.5$ & 42.7\% \\
Philadelphia & 11--11 & 41.0 & 41.0 & 46 & $-5.0$ & 26.8\% \\
Boston & 13--17 & 35.5 & 36.9 & 42 & $-6.5$ & 36.6\% \\
Washington & 4--7 & 29.8 & 35.1 & 40 & $-10.2$ & 13.4\% \\
New Jersey & 2--4 & 27.3 & 35.9 & 26 & $+1.3$ & 7.3\% \\
Charlotte & 3--10 & 18.9 & 28.5 & 20 & $-1.1$ & 15.9\% \\
Detroit & 25--7 & 64.1 & 58.6 & 63 & $+1.1$ & 39.0\% \\
Cleveland & 11--8 & 47.5 & 45.2 & 57 & $-9.5$ & 23.2\% \\
Atlanta & 11--12 & 39.2 & 39.8 & 52 & $-12.8$ & 28.0\% \\
Milwaukee & 9--5 & 52.7 & 47.8 & 49 & $+3.7$ & 17.1\% \\
Chicago & 23--19 & 44.9 & 44.2 & 47 & $-2.1$ & 51.2\% \\
Indiana & 3--13 & 15.4 & 25.2 & 28 & $-12.6$ & 19.5\% \\
Utah & 6--7 & 37.8 & 39.2 & 51 & $-13.2$ & 15.9\% \\
Houston & 6--8 & 35.1 & 37.6 & 45 & $-9.9$ & 17.1\% \\
Denver & 6--9 & 32.8 & 36.1 & 44 & $-11.2$ & 18.3\% \\
Dallas & 3--8 & 22.4 & 31.2 & 38 & $-15.6$ & 13.4\% \\
San Antonio & 1--4 & 16.4 & 32.8 & 21 & $-4.6$ & 6.1\% \\
Miami & 2--9 & 14.9 & 27.3 & 15 & $-0.1$ & 13.4\% \\
LA Lakers & 52--25 & 55.4 & 53.7 & 57 & $-1.6$ & 93.9\% \\
Phoenix & 11--7 & 50.1 & 46.9 & 55 & $-4.9$ & 22.0\% \\
Seattle & 10--6 & 51.2 & 47.3 & 47 & $+4.2$ & 19.5\% \\
Golden State & 7--6 & 44.2 & 42.8 & 43 & $+1.2$ & 15.9\% \\
Portland & 8--14 & 29.8 & 33.3 & 39 & $-9.2$ & 26.8\% \\
Sacramento & 2--5 & 23.4 & 33.8 & 27 & $-3.6$ & 8.5\% \\
LA Clippers & 3--8 & 22.4 & 31.2 & 21 & $+1.4$ & 13.4\% \\
\midrule
\textbf{League MAE} & & \textbf{6.4} & \textbf{7.0} & --- & --- & --- \\
\bottomrule
\end{tabular}
\end{table}

\subsection*{1989--90 Season}
\textit{249 games logged (22.5\% coverage). $\lambda = 1{,}125$.}

\begin{table}[h]
\centering
\small
\caption{Win-loss validation, 1989--90 NBA season.}
\label{tab:val_8990}
\begin{tabular}{lrrrrrrr}
\toprule
Franchise & Sampled & MLE & Bayes & Truth & Error & \% Samp \\
\midrule
Philadelphia & 10--6 & 51.2 & 47.3 & 53 & $-1.8$ & 19.5\% \\
Boston & 29--12 & 58.0 & 54.7 & 52 & $+6.0$ & 50.0\% \\
New York & 10--14 & 34.2 & 36.2 & 45 & $-10.8$ & 29.3\% \\
Washington & 2--4 & 27.3 & 35.9 & 31 & $-3.7$ & 7.3\% \\
Miami & 0--7 & 0.0 & 24.1 & 18 & $-18.0$ & 8.5\% \\
New Jersey & 2--10 & 13.7 & 26.1 & 17 & $-3.3$ & 14.6\% \\
Detroit & 15--8 & 53.5 & 49.7 & 59 & $-5.5$ & 28.0\% \\
Chicago & 36--20 & 52.7 & 50.9 & 55 & $-2.3$ & 68.3\% \\
Milwaukee & 3--9 & 20.5 & 29.8 & 44 & $-23.5$ & 14.6\% \\
Cleveland & 3--14 & 14.5 & 24.3 & 42 & $-27.5$ & 20.7\% \\
Indiana & 6--9 & 32.8 & 36.1 & 42 & $-9.2$ & 18.3\% \\
Atlanta & 5--12 & 24.1 & 30.4 & 41 & $-16.9$ & 20.7\% \\
Orlando & 5--5 & 41.0 & 41.0 & 18 & $+23.0$ & 12.2\% \\
San Antonio & 12--12 & 41.0 & 41.0 & 56 & $-15.0$ & 29.3\% \\
Utah & 5--8 & 31.5 & 35.7 & 55 & $-23.5$ & 15.9\% \\
Dallas & 4--10 & 23.4 & 30.8 & 47 & $-23.6$ & 17.1\% \\
Denver & 3--11 & 17.6 & 27.3 & 43 & $-25.4$ & 17.1\% \\
Houston & 9--7 & 46.1 & 44.2 & 41 & $+5.1$ & 19.5\% \\
Minnesota & 2--5 & 23.4 & 33.8 & 22 & $+1.4$ & 8.5\% \\
Charlotte & 0--9 & 0.0 & 21.6 & 19 & $-19.0$ & 11.0\% \\
LA Lakers & 47--11 & 66.4 & 62.7 & 63 & $+3.4$ & 70.7\% \\
Portland & 15--5 & 61.5 & 54.7 & 59 & $+2.5$ & 24.4\% \\
Phoenix & 10--6 & 51.2 & 47.3 & 54 & $-2.8$ & 19.5\% \\
Seattle & 5--12 & 24.1 & 30.4 & 41 & $-16.9$ & 20.7\% \\
Golden State & 8--9 & 38.6 & 39.5 & 37 & $+1.6$ & 20.7\% \\
LA Clippers & 2--5 & 23.4 & 33.8 & 30 & $-6.6$ & 8.5\% \\
Sacramento & 1--9 & 8.2 & 24.6 & 23 & $-14.8$ & 12.2\% \\
\midrule
\textbf{League MAE} & & \textbf{11.6} & \textbf{8.7} & --- & --- & --- \\
\bottomrule
\end{tabular}
\end{table}

\subsection*{1990--91 Season}
\textit{297 games logged (26.8\% coverage). $\lambda = 1{,}341$.}

\begin{table}[h]
\centering
\small
\caption{Win-loss validation, 1990--91 NBA season.}
\label{tab:val_9091}
\begin{tabular}{lrrrrrrr}
\toprule
Franchise & Sampled & MLE & Bayes & Truth & Error & \% Samp \\
\midrule
Boston & 30--11 & 60.0 & 56.3 & 56 & $+4.0$ & 50.0\% \\
Philadelphia & 12--11 & 42.8 & 42.2 & 44 & $-1.2$ & 28.0\% \\
New York & 11--14 & 36.1 & 37.5 & 39 & $-2.9$ & 30.5\% \\
New Jersey & 4--10 & 23.4 & 30.8 & 26 & $-2.6$ & 17.1\% \\
Miami & 2--7 & 18.2 & 30.2 & 24 & $-5.8$ & 11.0\% \\
Washington & 3--10 & 18.9 & 28.5 & 30 & $-11.1$ & 15.9\% \\
Chicago & 37--20 & 53.2 & 51.4 & 61 & $-7.8$ & 69.5\% \\
Detroit & 25--21 & 44.6 & 43.9 & 50 & $-5.4$ & 56.1\% \\
Atlanta & 14--15 & 39.6 & 39.9 & 43 & $-3.4$ & 35.4\% \\
Milwaukee & 5--8 & 31.5 & 35.7 & 48 & $-16.5$ & 15.9\% \\
Indiana & 5--8 & 31.5 & 35.7 & 41 & $-9.5$ & 15.9\% \\
Charlotte & 2--9 & 14.9 & 27.3 & 26 & $-11.1$ & 13.4\% \\
Cleveland & 3--13 & 15.4 & 25.2 & 33 & $-17.6$ & 19.5\% \\
San Antonio & 34--20 & 51.6 & 50.0 & 55 & $-3.4$ & 65.9\% \\
Houston & 15--10 & 49.2 & 46.9 & 52 & $-2.8$ & 30.5\% \\
Utah & 9--11 & 36.9 & 38.3 & 54 & $-17.1$ & 24.4\% \\
Dallas & 3--6 & 27.3 & 34.5 & 28 & $-0.7$ & 11.0\% \\
Orlando & 2--6 & 20.5 & 31.9 & 31 & $-10.5$ & 9.8\% \\
Minnesota & 1--3 & 20.5 & 35.1 & 29 & $-8.5$ & 4.9\% \\
Denver & 5--22 & 15.2 & 22.2 & 20 & $-4.8$ & 32.9\% \\
LA Lakers & 32--19 & 51.5 & 49.7 & 58 & $-6.5$ & 62.2\% \\
Phoenix & 14--7 & 54.7 & 50.3 & 55 & $-0.3$ & 25.6\% \\
Portland & 12--8 & 49.2 & 46.5 & 63 & $-13.8$ & 24.4\% \\
Golden State & 8--8 & 41.0 & 41.0 & 44 & $-3.0$ & 19.5\% \\
Seattle & 9--14 & 32.1 & 34.8 & 41 & $-8.9$ & 28.0\% \\
Sacramento & 0--2 & 0.0 & 34.2 & 25 & $-25.0$ & 2.4\% \\
LA Clippers & 0--4 & 0.0 & 29.3 & 31 & $-31.0$ & 4.9\% \\
\midrule
\textbf{League MAE} & & \textbf{8.7} & \textbf{5.7} & --- & --- & --- \\
\bottomrule
\end{tabular}
\end{table}

\subsection*{1991--92 Season}
\textit{249 games logged (22.5\% coverage). $\lambda = 1{,}125$.}

\begin{table}[h]
\centering
\small
\caption{Win-loss validation, 1991--92 NBA season.}
\label{tab:val_9192}
\begin{tabular}{lrrrrrrr}
\toprule
Franchise & Sampled & MLE & Bayes & Truth & Error & \% Samp \\
\midrule
Boston & 24--11 & 56.2 & 52.8 & 51 & $+5.2$ & 42.7\% \\
New York & 24--18 & 46.9 & 45.7 & 51 & $-4.1$ & 51.2\% \\
New Jersey & 8--10 & 36.4 & 38.1 & 40 & $-3.6$ & 22.0\% \\
Miami & 2--9 & 14.9 & 27.3 & 38 & $-23.1$ & 13.4\% \\
Philadelphia & 4--14 & 18.2 & 26.4 & 35 & $-16.8$ & 22.0\% \\
Washington & 3--10 & 18.9 & 28.5 & 25 & $-6.1$ & 15.9\% \\
Orlando & 1--11 & 6.8 & 22.4 & 21 & $-14.2$ & 14.6\% \\
Chicago & 57--14 & 65.8 & 62.8 & 67 & $-1.2$ & 86.6\% \\
Cleveland & 2--6 & 20.5 & 31.9 & 57 & $-36.5$ & 9.8\% \\
Detroit & 18--9 & 54.7 & 51.0 & 48 & $+6.7$ & 32.9\% \\
Indiana & 5--7 & 34.2 & 37.3 & 40 & $-5.8$ & 14.6\% \\
Atlanta & 7--11 & 31.9 & 35.1 & 38 & $-6.1$ & 22.0\% \\
Milwaukee & 4--8 & 27.3 & 33.5 & 31 & $-3.7$ & 14.6\% \\
Charlotte & 2--9 & 14.9 & 27.3 & 31 & $-16.1$ & 13.4\% \\
Utah & 6--7 & 37.8 & 39.2 & 55 & $-17.2$ & 15.9\% \\
San Antonio & 9--9 & 41.0 & 41.0 & 47 & $-6.0$ & 22.0\% \\
Houston & 10--8 & 45.6 & 43.9 & 42 & $+3.6$ & 22.0\% \\
Denver & 3--9 & 20.5 & 29.8 & 24 & $-3.5$ & 14.6\% \\
Dallas & 1--8 & 9.1 & 25.9 & 22 & $-12.9$ & 11.0\% \\
Minnesota & 1--5 & 13.7 & 30.8 & 15 & $-1.3$ & 7.3\% \\
Portland & 25--15 & 51.2 & 49.2 & 57 & $-5.8$ & 48.8\% \\
Golden State & 7--8 & 38.3 & 39.4 & 55 & $-16.7$ & 18.3\% \\
Phoenix & 10--3 & 63.1 & 53.5 & 53 & $+10.1$ & 15.9\% \\
Seattle & 3--7 & 24.6 & 32.8 & 47 & $-22.4$ & 12.2\% \\
LA Clippers & 2--7 & 18.2 & 30.2 & 45 & $-26.8$ & 11.0\% \\
LA Lakers & 9--9 & 41.0 & 41.0 & 43 & $-2.0$ & 22.0\% \\
Sacramento & 2--7 & 18.2 & 30.2 & 29 & $-10.8$ & 11.0\% \\
\midrule
\textbf{League MAE} & & \textbf{10.7} & \textbf{6.8} & --- & --- & --- \\
\bottomrule
\end{tabular}
\end{table}

\subsection*{1992--93 Season}
\textit{301 games logged (27.2\% coverage). $\lambda = 1{,}360$.}

\begin{table}[h]
\centering
\small
\caption{Win-loss validation, 1992--93 NBA season.}
\label{tab:val_9293}
\begin{tabular}{lrrrrrrr}
\toprule
Franchise & Sampled & MLE & Bayes & Truth & Error & \% Samp \\
\midrule
New York & 36--15 & 57.9 & 55.1 & 60 & $-2.1$ & 62.2\% \\
Boston & 13--10 & 46.3 & 44.7 & 48 & $-1.7$ & 28.0\% \\
New Jersey & 19--12 & 50.3 & 48.0 & 43 & $+7.3$ & 37.8\% \\
Orlando & 17--16 & 42.2 & 42.0 & 41 & $+1.2$ & 40.2\% \\
Miami & 4--9 & 25.2 & 32.1 & 36 & $-10.8$ & 15.9\% \\
Philadelphia & 5--13 & 22.8 & 29.3 & 26 & $-3.2$ & 22.0\% \\
Washington & 1--12 & 6.3 & 21.4 & 22 & $-15.7$ & 15.9\% \\
Chicago & 54--25 & 56.1 & 54.4 & 57 & $-0.9$ & 96.3\% \\
Cleveland & 8--5 & 50.5 & 46.3 & 54 & $-3.5$ & 15.9\% \\
Charlotte & 10--13 & 35.7 & 37.3 & 44 & $-8.3$ & 28.0\% \\
Atlanta & 17--11 & 49.8 & 47.5 & 43 & $+6.8$ & 34.1\% \\
Indiana & 8--10 & 36.4 & 38.1 & 41 & $-4.6$ & 22.0\% \\
Detroit & 13--21 & 31.4 & 33.5 & 40 & $-8.6$ & 41.5\% \\
Milwaukee & 2--11 & 12.6 & 25.0 & 28 & $-15.4$ & 15.9\% \\
Houston & 9--6 & 49.2 & 45.9 & 55 & $-5.8$ & 18.3\% \\
San Antonio & 10--13 & 35.7 & 37.3 & 49 & $-13.3$ & 28.0\% \\
Utah & 3--7 & 24.6 & 32.8 & 47 & $-22.4$ & 12.2\% \\
Denver & 5--5 & 41.0 & 41.0 & 36 & $+5.0$ & 12.2\% \\
Minnesota & 5--12 & 24.1 & 30.4 & 19 & $+5.1$ & 20.7\% \\
Dallas & 2--6 & 20.5 & 31.9 & 11 & $+9.5$ & 9.8\% \\
Phoenix & 17--7 & 58.1 & 53.1 & 62 & $-3.9$ & 29.3\% \\
Seattle & 14--9 & 49.9 & 47.2 & 55 & $-5.1$ & 28.0\% \\
Portland & 11--11 & 41.0 & 41.0 & 51 & $-10.0$ & 26.8\% \\
LA Clippers & 5--8 & 31.5 & 35.7 & 41 & $-9.5$ & 15.9\% \\
LA Lakers & 10--10 & 41.0 & 41.0 & 39 & $+2.0$ & 24.4\% \\
Golden State & 2--16 & 9.1 & 20.5 & 34 & $-24.9$ & 22.0\% \\
Sacramento & 1--8 & 9.1 & 25.9 & 25 & $-15.9$ & 11.0\% \\
\midrule
\textbf{League MAE} & & \textbf{8.2} & \textbf{6.5} & --- & --- & --- \\
\bottomrule
\end{tabular}
\end{table}

\subsection*{1993--94 Season}
\textit{20 games logged (1.8\% coverage). $\lambda = 90$.\\ \textbf{Caution: stub season (20 games). Estimates unreliable.}}

\begin{table}[h]
\centering
\small
\caption{Win-loss validation, 1993--94 NBA season.}
\label{tab:val_9394}
\begin{tabular}{lrrrrrrr}
\toprule
Franchise & Sampled & MLE & Bayes & Truth & Error & \% Samp \\
\midrule
New York & 3--0 & 82.0 & 50.5 & 57 & $+25.0$ & 3.7\% \\
Orlando & 1--0 & 82.0 & 44.7 & 50 & $+32.0$ & 1.2\% \\
Miami & 1--0 & 82.0 & 44.7 & 42 & $+40.0$ & 1.2\% \\
Boston & 1--2 & 27.3 & 37.8 & 32 & $-4.7$ & 3.7\% \\
Philadelphia & 1--1 & 41.0 & 41.0 & 25 & $+16.0$ & 2.4\% \\
Atlanta & 3--1 & 61.5 & 46.9 & 57 & $+4.5$ & 4.9\% \\
Chicago & 2--1 & 54.7 & 44.2 & 55 & $-0.3$ & 3.7\% \\
Cleveland & 2--0 & 82.0 & 47.8 & 47 & $+35.0$ & 2.4\% \\
Charlotte & 0--1 & 0.0 & 37.3 & 41 & $-41.0$ & 1.2\% \\
Indiana & 0--1 & 0.0 & 37.3 & 47 & $-47.0$ & 1.2\% \\
Houston & 0--1 & 0.0 & 37.3 & 58 & $-58.0$ & 1.2\% \\
San Antonio & 1--0 & 82.0 & 44.7 & 55 & $+27.0$ & 1.2\% \\
Utah & 0--1 & 0.0 & 37.3 & 53 & $-53.0$ & 1.2\% \\
Denver & 0--2 & 0.0 & 34.2 & 42 & $-42.0$ & 2.4\% \\
Minnesota & 1--1 & 41.0 & 41.0 & 20 & $+21.0$ & 2.4\% \\
Dallas & 0--2 & 0.0 & 34.2 & 13 & $-13.0$ & 2.4\% \\
Seattle & 1--0 & 82.0 & 44.7 & 63 & $+19.0$ & 1.2\% \\
Phoenix & 1--1 & 41.0 & 41.0 & 56 & $-15.0$ & 2.4\% \\
Golden State & 0--1 & 0.0 & 37.3 & 50 & $-50.0$ & 1.2\% \\
Portland & 0--2 & 0.0 & 34.2 & 47 & $-47.0$ & 2.4\% \\
LA Lakers & 0--1 & 0.0 & 37.3 & 33 & $-33.0$ & 1.2\% \\
Sacramento & 1--1 & 41.0 & 41.0 & 28 & $+13.0$ & 2.4\% \\
LA Clippers & 1--0 & 82.0 & 44.7 & 27 & $+55.0$ & 1.2\% \\
\midrule
\textbf{League MAE} & & \textbf{30.1} & \textbf{11.4} & --- & --- & --- \\
\bottomrule
\end{tabular}
\end{table}

\subsection*{1994--95 Season}
\textit{10 games logged (0.9\% coverage). $\lambda = 45$.\\ \textbf{Caution: stub season (10 games). Estimates unreliable.}}

\begin{table}[h]
\centering
\small
\caption{Win-loss validation, 1994--95 NBA season.}
\label{tab:val_9495}
\begin{tabular}{lrrrrrrr}
\toprule
Franchise & Sampled & MLE & Bayes & Truth & Error & \% Samp \\
\midrule
New York & 2--0 & 82.0 & 47.8 & 55 & $+27.0$ & 2.4\% \\
Boston & 0--1 & 0.0 & 37.3 & 35 & $-35.0$ & 1.2\% \\
Philadelphia & 0--2 & 0.0 & 34.2 & 24 & $-24.0$ & 2.4\% \\
Washington & 0--1 & 0.0 & 37.3 & 21 & $-21.0$ & 1.2\% \\
Chicago & 2--0 & 82.0 & 47.8 & 47 & $+35.0$ & 2.4\% \\
Atlanta & 0--1 & 0.0 & 37.3 & 42 & $-42.0$ & 1.2\% \\
Detroit & 1--0 & 82.0 & 44.7 & 28 & $+54.0$ & 1.2\% \\
San Antonio & 0--1 & 0.0 & 37.3 & 62 & $-62.0$ & 1.2\% \\
Denver & 1--0 & 82.0 & 44.7 & 41 & $+41.0$ & 1.2\% \\
Minnesota & 1--1 & 41.0 & 41.0 & 21 & $+20.0$ & 2.4\% \\
Phoenix & 2--0 & 82.0 & 47.8 & 59 & $+23.0$ & 2.4\% \\
Seattle & 0--1 & 0.0 & 37.3 & 57 & $-57.0$ & 1.2\% \\
Portland & 0--1 & 0.0 & 37.3 & 44 & $-44.0$ & 1.2\% \\
Sacramento & 1--0 & 82.0 & 44.7 & 39 & $+43.0$ & 1.2\% \\
Golden State & 0--1 & 0.0 & 37.3 & 26 & $-26.0$ & 1.2\% \\
\midrule
\textbf{League MAE} & & \textbf{36.9} & \textbf{10.7} & --- & --- & --- \\
\bottomrule
\end{tabular}
\end{table}

\subsection*{1995--96 Season}
\textit{264 games logged (22.2\% coverage). $\lambda = 1{,}110$.}

\begin{table}[h]
\centering
\small
\caption{Win-loss validation, 1995--96 NBA season.}
\label{tab:val_9596}
\begin{tabular}{lrrrrrrr}
\toprule
Franchise & Sampled & MLE & Bayes & Truth & Error & \% Samp \\
\midrule
Orlando & 24--6 & 65.6 & 59.4 & 60 & $+5.6$ & 36.6\% \\
New York & 18--17 & 42.2 & 41.9 & 47 & $-4.8$ & 42.7\% \\
Miami & 4--11 & 21.9 & 29.5 & 42 & $-20.1$ & 18.3\% \\
Washington & 3--8 & 22.4 & 31.2 & 39 & $-16.6$ & 13.4\% \\
Boston & 3--8 & 22.4 & 31.2 & 33 & $-10.6$ & 13.4\% \\
New Jersey & 4--8 & 27.3 & 33.5 & 30 & $-2.7$ & 14.6\% \\
Philadelphia & 3--14 & 14.5 & 24.3 & 18 & $-3.5$ & 20.7\% \\
Chicago & 72--10 & 72.0 & 68.6 & 72 & $0.0$ & 100.0\% \\
Indiana & 6--8 & 35.1 & 37.6 & 52 & $-16.9$ & 17.1\% \\
Cleveland & 3--7 & 24.6 & 32.8 & 47 & $-22.4$ & 12.2\% \\
Detroit & 7--8 & 38.3 & 39.4 & 46 & $-7.7$ & 18.3\% \\
Atlanta & 3--9 & 20.5 & 29.8 & 46 & $-25.5$ & 14.6\% \\
Charlotte & 10--9 & 43.2 & 42.4 & 41 & $+2.2$ & 23.2\% \\
Toronto & 5--9 & 29.3 & 34.2 & 21 & $+8.3$ & 17.1\% \\
Milwaukee & 2--9 & 14.9 & 27.3 & 25 & $-10.1$ & 13.4\% \\
San Antonio & 8--5 & 50.5 & 46.3 & 59 & $-8.5$ & 15.9\% \\
Utah & 4--8 & 27.3 & 33.5 & 55 & $-27.7$ & 14.6\% \\
Houston & 9--11 & 36.9 & 38.3 & 48 & $-11.1$ & 24.4\% \\
Denver & 5--8 & 31.5 & 35.7 & 35 & $-3.5$ & 15.9\% \\
Dallas & 7--9 & 35.9 & 37.8 & 26 & $+9.9$ & 19.5\% \\
Minnesota & 4--13 & 19.3 & 27.3 & 26 & $-6.7$ & 20.7\% \\
Vancouver & 3--9 & 20.5 & 29.8 & 15 & $+5.5$ & 14.6\% \\
Seattle & 15--8 & 53.5 & 49.7 & 64 & $-10.5$ & 28.0\% \\
LA Lakers & 18--10 & 52.7 & 49.6 & 53 & $-0.3$ & 34.1\% \\
Portland & 7--5 & 47.8 & 44.7 & 44 & $+3.8$ & 14.6\% \\
Phoenix & 5--13 & 22.8 & 29.3 & 41 & $-18.2$ & 22.0\% \\
Sacramento & 4--8 & 27.3 & 33.5 & 39 & $-11.7$ & 14.6\% \\
Golden State & 5--11 & 25.6 & 31.5 & 36 & $-10.4$ & 19.5\% \\
LA Clippers & 3--5 & 30.8 & 36.4 & 29 & $+1.8$ & 9.8\% \\
\midrule
\textbf{League MAE} & & \textbf{9.9} & \textbf{7.9} & --- & --- & --- \\
\bottomrule
\end{tabular}
\end{table}

\section{Annotated RAPM Estimation Code}
\label{app:code}

The following Python script implements the RAPM estimation procedure
described in Section~\ref{sec:estimation}. The script is invoked from
the command line as \texttt{python rapm.py <season>}, where
\texttt{<season>} is a two-digit integer corresponding to the final
year of the NBA season (e.g., \texttt{91} for the 1990--91 season).
The stint data file is expected at the path
\texttt{HistoricalStints<season>.csv}.

\begin{small}
\begin{verbatim}
import sys
import numpy as np
import pandas as pd
import matplotlib.pyplot as plt
from scipy.stats import iqr
from sklearn.neighbors import KernelDensity

# -----------------------------------------------------------
# 1. Load stint data
#    Columns: Oteam, Dteam, O1-O5 (offensive lineup),
#    D1-D5 (defensive lineup), Oposs, Dposs, Oscore, Dscore
# -----------------------------------------------------------
season = str(sys.argv[1])
data = pd.read_csv('HistoricalStints' + season + '.csv')
dataColumns = data.columns

# -----------------------------------------------------------
# 2. Player enumeration and possession accumulation
#    Iterate over ALL rows (even + odd) to build the full
#    player roster and accumulate offensive/defensive
#    possession and point totals per player.
#    Only even-indexed rows are used in the regression
#    (odd rows are the mirror-image perspective of each
#    game stint and are excluded to avoid double-counting).
# -----------------------------------------------------------
players = {}       # player name -> integer index
playerTeam = {}    # player name -> team abbreviation
playerPoss = {}    # player name -> [off_poss, def_poss]
playerPoints = {}  # player name -> [off_pts, def_pts]
indexNumber = 0
zeroPoss = 0       # count of stints with Oposs < 1
maxStint = 0       # maximum possession count in any stint

for index, row in data.iterrows():
    counter = 0
    # Columns 4-13 are O1-O5, D1-D5
    for col in dataColumns[4:14]:
        counter += 1
        if row[col] not in players:
            players[row[col]] = indexNumber
            # Columns 4-8 (counter 1-5): offensive players
            if counter < 6:
                playerTeam[row[col]] = row['Oteam']
                playerPoss[row[col]] = [0, 0]
                playerPoints[row[col]] = [0, 0]
            # Columns 9-13 (counter 6-10): defensive players
            else:
                playerTeam[row[col]] = row['Dteam']
                playerPoss[row[col]] = [0, 0]
                playerPoints[row[col]] = [0, 0]
            indexNumber += 1
    if float(row['Oposs']) < 1:
        zeroPoss += 1

# Accumulate possession/point totals on even rows only
for index, row in data.iterrows():
    if (index % 2) == 0:
        counter = 0
        # Track the maximum single-stint possession count
        if row['Oposs'] > maxStint:
            maxStint = row['Oposs']
        elif row['Dposs'] > maxStint:
            maxStint = row['Dposs']
        for col in dataColumns[4:14]:
            counter += 1
            if counter < 6:
                # Offensive player: Oposs = offensive poss,
                #                   Dposs = defensive poss
                playerPoss[row[col]][0] += int(row['Oposs'])
                playerPoss[row[col]][1] += int(row['Dposs'])
                playerPoints[row[col]][0] += int(row['Oscore'])
                playerPoints[row[col]][1] += int(row['Dscore'])
            else:
                # Defensive player: roles are reversed
                playerPoss[row[col]][1] += int(row['Oposs'])
                playerPoss[row[col]][0] += int(row['Dposs'])
                playerPoints[row[col]][1] += int(row['Oscore'])
                playerPoints[row[col]][0] += int(row['Dscore'])

# -----------------------------------------------------------
# 3. Design matrix construction
#    Shape: (N - zeroPoss) x (2*P + 1)
#    Column 0:         intercept (all ones)
#    Columns 1..P:     offensive player indicators (+1)
#    Columns P+1..2P:  defensive player indicators (-1)
#    Response:         Oscore / Oposs * 100 (pts per 100 poss)
#    Weight:           Oposs
# -----------------------------------------------------------
numPlayers = len(players)
N_eff = len(data.index) - zeroPoss

stints  = np.zeros((N_eff, 2 * numPlayers + 1))
weights = np.zeros((N_eff, N_eff))
ratings = np.zeros((N_eff, 1))
offset  = 0

for index, row in data.iterrows():
    if float(row['Oposs']) > 0:
        stints[index - offset, 0] = 1  # intercept
        # Offensive players: columns 4-8 -> +1 in block 1
        for col in dataColumns[4:9]:
            stints[index - offset,
                   players[row[col]] + 1] = 1
        # Defensive players: columns 9-13 -> -1 in block 2
        for col in dataColumns[9:14]:
            stints[index - offset,
                   numPlayers + players[row[col]] + 1] = -1
        # Possession weight
        weights[index - offset,
                index - offset] = float(row['Oposs'])
        # Response: offensive points per 100 possessions
        ratings[index - offset] = (
            100. * float(row['Oscore']) / float(row['Oposs'])
        )
    else:
        offset += 1  # skip zero-possession stints

# -----------------------------------------------------------
# 4. Regularization parameter (coverage-scaled)
#    Formula: lambda = (G_logged / G_season) * 5000
#    Update numerator as reconstruction progresses.
# -----------------------------------------------------------
# Example: 240 logged games, 1025-game season
weight = (240.00 / 1025.) * 5000.
# Other seasons (update numerator as coverage grows):
# weight = (142.00 / 1107.) * 5000.
# weight = (105.00 / 1189.) * 5000.

# -----------------------------------------------------------
# 5. Ridge regression
#    beta = (X'WX + lambda*I)^{-1} X'W y
# -----------------------------------------------------------
playerMatrix  = np.matrix(stints)
W             = np.matrix(weights)
ratingVector  = np.matrix(ratings)

beta = (
    np.linalg.inv(
        playerMatrix.T.dot(W).dot(playerMatrix)
        + weight * np.eye(1 + 2 * numPlayers)
    )
    .dot(playerMatrix.T.dot(W).dot(ratingVector))
)

# -----------------------------------------------------------
# 6. Residual variance estimation
#    sigma^2 = (N/2 - 2P - 1)^{-1} (y - X*beta)' W (y - X*beta)
#    N/2: effective observations (even rows only)
#    2P+1: number of parameters (intercept + P off + P def)
# -----------------------------------------------------------
sigmahat = (
    (data.shape[0] / 2 - 2 * numPlayers - 1) ** (-1)
    * (ratingVector - playerMatrix.dot(beta)).T
    .dot(W)
    .dot(ratingVector - playerMatrix.dot(beta))
)

# -----------------------------------------------------------
# 7. Posterior covariance matrix
#    Sigma = sigma^2 * (X'WX + lambda*I)^{-1}
# -----------------------------------------------------------
rapmerr = (
    sigmahat[0, 0]
    * np.linalg.inv(
        playerMatrix.T.dot(W).dot(playerMatrix)
        + weight * np.eye(1 + 2 * numPlayers)
    )
)

# -----------------------------------------------------------
# 8. RAPM extraction, mean-centering, and credible intervals
#    For player j:
#      raw_off  = beta[idx(j) + 1]
#      raw_def  = beta[P + idx(j) + 1]
#      stderr   = 1.96 * sqrt(Sigma[k1,k1] + Sigma[k2,k2])
#    ORAPM = raw_off - mean(all raw_off)
#    DRAPM = raw_def - mean(all raw_def)
#    RAPM  = ORAPM + DRAPM
# -----------------------------------------------------------
rapms = {}   # player -> raw total RAPM (pre-centering)
perrs = {}   # player -> [RAPM_low, RAPM_high]

for player in players:
    rapms[player] = (
        beta[players[player] + 1]
        + beta[players[player] + 1 + numPlayers]
    )
    k1     = players[player] + 1
    k2     = players[player] + 1 + numPlayers
    stderr = 1.96 * np.sqrt(rapmerr[k1, k1]
                            + rapmerr[k2, k2])
    perrs[player] = [
        rapms[player].tolist()[0][0] - stderr,
        rapms[player].tolist()[0][0] + stderr
    ]

# Sort by raw RAPM descending, compute mean offsets
sortRaps = sorted(rapms.items(), key=lambda kv: kv[1],
                  reverse=True)
aves = [0., 0.]
for player in sortRaps:
    aves[0] += beta[players[player[0]] + 1][0, 0]
    aves[1] += beta[players[player[0]] + 1 + numPlayers][0, 0]
aves[0] /= len(players)  # mean offensive coefficient
aves[1] /= len(players)  # mean defensive coefficient

# Print ranked output with centered ORAPM, DRAPM, RAPM
template = ("{0:5} {1:25} {2:5}|{3:7} {4:7}"
            "|{5:7} {6:7} | {7:7} {8:7} | {9:7} {10:7} {11:7}")
print(template.format('Rank', 'Player', 'Team',
      'OPoss', 'O PTS', 'DPoss', 'D PTS',
      'ORAPM', 'DRAPM', 'TOTAL', 'LOW', 'HIGH'))
rank = 1
for player in sortRaps:
    orapm = beta[players[player[0]] + 1][0, 0]     - aves[0]
    drapm = beta[players[player[0]] + 1 + numPlayers][0, 0] \
            - aves[1]
    print(template.format(
        rank, player[0],
        playerTeam[player[0]],
        playerPoss[player[0]][0],
        playerPoints[player[0]][0],
        playerPoss[player[0]][1],
        playerPoints[player[0]][1],
        '%.2f' % orapm,
        '%.2f' % drapm,
        '%.2f' % (rapms[player[0]][0, 0] - aves[0] - aves[1]),
        '%.2f' % perrs[player[0]][0],
        '%.2f' % perrs[player[0]][1]
    ))
    rank += 1
\end{verbatim}
\end{small}

\subsubsection*{Diagnostic output and visualizations}

After printing the ranked player table, the script produces three
diagnostic visualizations:

\begin{enumerate}[leftmargin=*, label=(\roman*)]
  \item \textbf{Stint-level scoring distribution}: A histogram of
    $y_i$ (points per 100 possessions) across all stints, with bin
    width set by the Freedman--Diaconis rule
    ($h = 2 \cdot \text{IQR}(y) / N^{1/3}$). This plot characterizes
    the distribution of scoring outcomes that the regression is
    fitting, and informs the choice of $\hat{\sigma}$.

  \item \textbf{Possession count distribution}: A histogram of
    $\texttt{Oposs}$ across all stints. This documents the distribution
    of stint lengths in the database and highlights the preponderance
    of short stints (1--3 possessions) that are possession-weighted
    down in the regression.

  \item \textbf{ORAPM and DRAPM density estimates}: Overlaid kernel
    density estimates of the centered offensive and defensive RAPM
    distributions, using Silverman's rule-of-thumb bandwidth
    ($h = 0.9 \cdot \min(\hat{\sigma}, \text{IQR}/1.34) \cdot n^{-1/5}$)
    applied separately to each distribution. This plot characterizes
    the spread of player quality in each dimension and provides a
    visual check for anomalous estimates.
\end{enumerate}

\end{document}